\documentclass[10pt,american]{article}
\usepackage[BOE]{express}
\usepackage{url}
\usepackage{caption}
\usepackage{relsize}

\usepackage[T1]{fontenc}
\usepackage{amsmath}
\usepackage{graphicx}

\usepackage[percent]{overpic}
\usepackage{cuted}

\def\Bias{\mathop{\operator@font Bias}\nolimits}
\def\E{\mathop{\operator@font E}\nolimits}
\def\P{\mathop{\operator@font P}\nolimits}
\def\Ci{\mathop{\operator@font Ci}\nolimits}
\def\Si{\mathop{\operator@font Si}\nolimits}
\def\Var{\mathop{\operator@font Var}\nolimits}
\def\sgn{\mathop{\operator@font sgn}\nolimits}

\begin{document}

\title{Plum pudding random medium model of biological tissue toward remote
microscopy from spectroscopic light scattering }

\author{Min Xu \authormark{1,*}} 
\address{\authormark{1} Department of Physics, Fairfield University,
  1073 North Benson Road, Fairfield, CT 06824, USA}
\email{\authormark{*}mxu@fairfield.edu}

\begin{abstract}
Biological tissue has a complex structure and exhibits rich spectroscopic
behavior. There is \emph{no} tissue model up to now able to account
for the observed spectroscopy of tissue light scattering and its anisotropy.
Here we present, \emph{for the first time}, a plum pudding random
medium (PPRM) model for biological tissue which succinctly describes
tissue as a superposition of distinctive scattering structures (plum)
embedded inside a fractal continuous medium of background refractive
index fluctuation (pudding). PPRM faithfully reproduces the wavelength
dependence of tissue light scattering and attributes the ``anomalous''
trend in the anisotropy to the plum and the powerlaw
dependence of the reduced scattering coefficient to the
fractal scattering pudding. Most importantly, PPRM opens
up a novel venue of quantifying the tissue
architecture and microscopic structures on average from macroscopic probing of the bulk with scattered light
alone without tissue excision. We demonstrate this potential by visualizing
the fine microscopic structural alterations in breast tissue (adipose,
glandular, fibrocystic, fibroadenoma, and ductal carcinoma) deduced
from noncontact spectroscopic measurement. 
\end{abstract}


\section{Introduction}

One central tenet in the application of light in biology and medicine
is noninvasive diagnosis of the structure and function of tissue from
tissue-light interaction \cite{Reviews1989}. Scattered light carries
important information about the morphology and optical properties
of the individual scatterers and can be used to identify structural
alterations or heterogeneities in tissue due to disease or physiological
variations \cite{yodh97:_diffus,Laughney2013,Wilson2015}. The scattering
and absorption properties of tissue determine light transport (such
as penetration, reflection, and transmission) and energy deposition
in tissue, key to both diagnostic and therapeutic applications of
light. Significant advances have been made during the past decades
in characterizing and modeling the optical properties of different
types of tissue (see, for example, \cite{cheong.ea90:_revie,Jacques2013},
for recent reviews). 

Biological tissue has a complex structure which determines the optical
properties of tissue. Microstructures in biological tissue range from
organelles $0.2-0.5\mu m$ or smaller, mitochondria $1-4\mu m$ in
length and $0.3-0.7\mu m$ in diameter, nuclei $3-10\mu m$ in diameter,
to mammalian cells $10-30\mu m$ in diameter. The refractive index
variation is about $0.04-0.10$ for soft tissue with a background
refractive index $n\simeq1.35-1.37$ \cite{schmittkumar98:_optical,xu07:_unified_mie_and_fract_scatt}.
When the wavelength, $\lambda$, of the probing light increases, light
is less scattered by tissue\cite{cheong.ea90:_revie,Jacques2013}
and the reduced scattering coefficient ($\mu_{s}'$) decreases. Light
is also expected to be more isotropically scattered into all directions
as the scatterers appear smaller with respect to the wavelength and
the anisotropy factor ($g$) defined as the mean cosine of the scattering
angles of tissue gets smaller. This widely-accepted notion about the
wavelength dependence of $\mu_{s}'$ and $g$ is, however, only partially
true and the trend of $g$ is contradictory to that found by thorough
measurements for various tissue types within visible and near-infrared
spectral range \cite{peters.ea90:_optic,Ma2005,Jacques2013}. The
``anomalous'' increase of $g$ with the probing wavelength seems
to be the rule rather than the exception for tissue light scattering
\cite{Jacques2013}.

Viewed on a microscopic scale, the constituents of tissue have no
clear boundaries and merge into a continuum structure. Furthermore,
many biological tissues have fractal-like organization and are statistically
self-similar \cite{and2.ea95:_scale-invariant,schmitt96:_turbul,and2.ea98:_self-affinity,vicsek01:_fluct}.
Light scattering property of a tissue was hence attributed to the
fluctuation of the refractive index distribution in tissue. A fractal
model \cite{schmitt96:_turbul,wang00:_model,xu05:_fract,Pu2012a}
and later a Whittle-Matern family of correlation functions \cite{sheppard2007,Rogers2014}
have been used successfully to describe such fluctuation in tissue.
These models show that $\mu_{s}'$ has a powerlaw dependence on the
wavelength ($\mu_{s}'\propto\lambda^{-b}$ with $b>0$ being the scattering
power) and correctly predict the decrease of tissue scattering with
$\lambda$ \cite{xu05:_fract}. Unfortunately, these models \emph{all}
predict the decrease of the anisotropy factor $g$ with $\lambda$,
disagreeing with experimental observations. The anisotropy factor
is one central parameter governing how light randomizes its directionality
and migrates with scattering in random media and bearing the direct
relation to the morphology and optical properties of the underlying
microscopic scattering constituents. The contradiction between experiments
and theoretical predictions on $g$ reveals the current lack in the
understanding of the nature of tissue light scattering. There is \emph{no}
tissue model up to now able to account for the observed spectroscopy
of scattering and its anisotropy. 

Recently, estimation of the effective scatterer size or nuclear morphology
in deep tissue from spectroscopic diffuse light measurements have
been reported \cite{wang2006,Hajihashemi2012}. This could potentially
lead to highly desirable \emph{in vivo} optical histopathology of
deep tissue from scattered light alone. An accurate yet succinct picture
and model of the complex structure of biological tissue will be the
foundation towards this direction, achieving extremely desirable remote
microscopy of biological tissue from bulk spectroscopic light scattering
without any tissue excision. 

In this article, after first reviewing continuum light scattering
models for tissue and identifying their deficiency. we present, \emph{for
the first time}, a Plum Pudding Random Medium model (PPRM) for biological
tissue. PPRM properly describes the scattering constituents in tissue
that tissue is a continuum yet with some prominent structures which
are distinctive from the background medium. In this unified view,
tissue light scattering is a superposition of both background refractive
index fluctuation and distinctive prominent structures. The distinctive
prominent structure is responsible for the observed ``anomalous''
anisotropy trend and provides a potential resolution to the long-lasting puzzle in the spectroscopic
properties of tissue. Afterwards, by establishing the explicit link
of the macroscopic scattering parameters of tissue to the microstructure,
we show PPRM opens up a new venue of quantifying
the \emph{microscopic} scattering constituents in tissue from \emph{macroscopic}
probing of a bulk from scattered light alone. We demonstrate this
potential at the end by visualizing the fine microscopic structural
alterations in breast tissue (normal adipose tissue, normal glandular
tissue, fibrocystic tissue, fibroadenoma, and ductal carcinoma) from
PPRM analysis of noncontact spectroscopic measurement.

\section{Theory}

\subsection{Background refractive index fluctuation}

One major source for light scattering by tissue is attributed to the
random fluctuation of the background refractive index for biological
tissues and cells \cite{xu05:_fract,xu06:_fract}. Denote $S_{\mathrm{bg}}(\mathbf{q})$
the scattering amplitude due to the random fluctuation of the background
refractive index. The squared background scattering amplitude is specified
by \cite{xu07:_unified_mie_and_fract_scatt}

\begingroup\medmuskip=0mu\thinmuskip=0mu\arraycolsep=1.4pt

\begin{equation}
\left|S_{\mathrm{bg}}(\mathbf{q})\right|^{2}\left\{ \begin{array}{c}
\mu^{2}\\
1
\end{array}\right.=2\pi k^{6}V\hat{R}(q)\left\{ \begin{array}{cc}
\mu^{2} & \mathrm{(parallel}\ \mathrm{polarized)}\\
1 & \mathrm{(perpendicular}\ \mathrm{polarized)}
\end{array}\right.\label{eq:S bg squared R(q)-1}
\end{equation}
\endgroup where $\mathbf{q}=q(\cos\phi,\sin\phi,0)$ is the wave
vector transfer with a magnitude $q=2k\sin\frac{\theta}{2}$, $k=2\pi n_{0}/\lambda$
is the wave number with $n_{0}$ the average refractive index of the
background medium and $\lambda$ the wavelength of the incident beam
in vacuum, $\theta,\;\phi$ are the polar and azimuthal angles of
scattering, respectively, $\mu\equiv\cos\theta$, $V$ is the volume,
and $\hat{R}(q)=\frac{1}{(2\pi)^{3}}\int R(r)\exp(i\mathbf{q}\cdot\mathbf{r})d\mathbf{r}$
is the power spectrum of the random fluctuation 
of the background refractive index specified by its correlation
function $R(\left|\mathbf{r}_{1}-\mathbf{r}_{2}\right|)=\left\langle
  \delta m(\mathbf{r}_{1})\delta m(\mathbf{r}_{2})\right\rangle $ with
$\left\langle \delta m(\mathbf{r})\right\rangle =0$.
The intensity of scattered parallel or perpendicular polarized light
is proportional to the corresponding scattering cross section given
by $\overline{\left|S_{\mathrm{bg}}(\mathbf{q})\right|^{2}}/k^{2}$
for light of respective polarization. The scattering cross section
for unpolarized light is given by the mean of the two scattering cross
sections for light of parallel or perpendicular polarization. The
differential scattering cross section for light scattering into the
direction $(\theta,\phi)$ from an incident beam linearly polarized
along the $x$ axis ($\phi=0$) is given by
\begin{equation}
\sigma(\theta,\phi)=2\pi k^{4}V\hat{R}(q)(\sin^{2}\phi+\mu^{2}\cos^{2}\phi).\label{eq:differential csca}
\end{equation}
The scattering and reduced scattering coefficients of the medium are
then expressed as

\begingroup\medmuskip=0mu\thickmuskip=0mu

\begin{equation}
\mu_{s\mathrm{,bg}}=\int\sigma(\theta,\phi)d\Omega=\frac{\pi}{k^{2}}\int_{-1}^{1}\left|S_{\mathrm{bg}}(q)\right|^{2}(1+\mu^{2})d\mu,\label{eq:csca}
\end{equation}

\begin{equation}
\mu_{s\mathrm{,bg}}'=\int\sigma(\theta,\phi)(1-\mu)d\Omega=\frac{\pi}{k^{2}}\int_{-1}^{1}\left|S_{\mathrm{bg}}(q)\right|^{2}(1+\mu^{2})(1-\mu)d\mu\label{eq:reduced csca}
\end{equation}
\endgroup after setting $V$ in Eq.~(\ref{eq:S bg squared R(q)-1})
to be unity. 

The fractal random continuous medium model \cite{xu05:_fract} assumes
the correlation function of
the random fluctuation of the background refractive index to be

\begin{equation}
R(r)=\beta^{2}\left(\frac{r}{l_{\mathrm{max}}}\right)^{4-D_{f}}\Gamma\left[-(4-D_{f}),\frac{r}{l_{\mathrm{max}}}\right].\label{eq:correlation R(r) fractal}
\end{equation}
Here the distribution of the correlation length $l$ is given by
$\eta(l)=\eta_{0}l^{3-D_{f}}/l_{\mathrm{max}}^{4-D_{f}}\quad(0\le l\le
l_{\mathrm{max}})$ normalized to $\int_{0}^{l_{\mathrm{max}}}\eta(l)dl=1$,
$\eta_{0}$ is a dimensionless constant, $\beta^{2}\equiv\left\langle \delta m(0)^{2}\right\rangle \eta_{0}$
represents the effective random fluctuation strength where
$\left\langle \delta m(0)^{2}\right\rangle $ is the squared amplitude 
fluctuation of the refractive index, $D_{f}$ is
the fractal dimension, and $\Gamma$ is the incomplete Gamma
function. For typical soft tissue, $0<D_f<7$ and $\sqrt{\left\langle \delta m(0)^{2}\right\rangle} \sim 0.01$.
The value of $\eta_{0}=4-D_{f}$ when $D_{f}<4$. The cutoff correlation
length $l_{\mathrm{max}}$ is the outer scale and $0$ is the inner
scale in Eq.~(\ref{eq:correlation R(r) fractal}). Strictly, when $D_{f}\ge4$,
the inner scale is no longer exactly $0$ and $\eta_{0}$ depends
also on the nonzero inner scale $l_{\mathrm{min}}$ ($\eta_{0}=(4-D_{f})/\left[1-\left(\frac{l_{\mathrm{min}}}{l_{\mathrm{max}}}\right)^{4-D_{f}}\right]$
for $D_{f}\neq4$ and $\eta_{0}=1/\log\left(\frac{l_{\mathrm{max}}}{l_{\mathrm{min}}}\right)$
for $D_{f}=4$). However, in this case Eq.~(\ref{eq:correlation R(r) fractal})
can still be used as light scattering by fluctuations of a correlation
length below the inner scale is much smaller and can be ignored. 
The background squared scattering amplitude is now
\begin{equation}
\left|S_{\mathrm{bg}}(\mathbf{q})\right|^{2}=\frac{2}{\pi}\frac{\beta^{2}Vk^{3}X^{3}}{7-D_{f}}{}_{2}F_{1}(2,\frac{7-D_{f}}{2},\frac{9-D_{f}}{2},-2(1-\mu)X^{2})\label{eq:S_bg squared}
\end{equation}
if $D_{f}<7$ where the size parameter $X\equiv kl_{\mathrm{max}}$
and $_{2}F_{1}$ is the Gauss hypergeometric function. The fractal
continuum medium model bears a close connection to a power law size
distribution of scatterers. Indeed, using the approximate amplitude
scattering matrix \cite{remizovich84:_theor} for spherical particles,
a discrete particle model assuming a particle size distribution of
the power law (number density of particles $\propto a^{-D_{f}}$ where
$a$ is the radius) shall yield the same amplitude scattering function
(\ref{eq:S_bg squared}) as in the fractal continuous medium model.
This illustrates the correlation length $l$ in the fractal continuous
medium model may be intuitively interpreted as the \emph{radius} of
``fictional'' scattering centers present within tissue \cite{xu06:_fract}.
The number density of the scattering centers of radius $l$ distributes
according to a power law $l^{-D_{f}}$. 

The Whittle-Matern family of correlation function for the fluctuation
of the background refractive index takes the form of 
\begin{equation}
R(\mathbf{r})=\left\langle \delta m(0)^{2}\right\rangle 2^{1-\nu}\left|\Gamma(\nu)\right|{}^{-1}\left(\frac{r}{l}\right)^{\nu}K_{\nu}\left(\frac{r}{l}\right)\label{eq:R(r) Whittle-Matern}
\end{equation}
where $K_{\nu}$ is the modified Bessel function of the second kind.
The Whittle-Matern correlation function has been used extensively
to model turbulence \cite{Uscinski1981} and was later used to model
tissue light scattering \cite{schmitt96:_turbul,sheppard2007,Rogers2014}.
 The parameter $l$ is the outer scale. When $-3/2<\nu<0$, Eq.~(\ref{eq:R(r) Whittle-Matern})
can still be used (with an implicit \emph{nonzero} inner scale). 
The Whittle-Matern random medium model gives:
\begin{equation}
\left|S_{\mathrm{bg}}(\mathbf{q})\right|^{2}=2\left\langle \delta m(0)^{2}\right\rangle \frac{\Gamma(\nu+\tfrac{3}{2})Vk^{3}X^{3}}{\pi^{1/2}\left|\Gamma(\nu)\right|}\left[1+2(1-\mu)X^{2}\right]^{-\nu-\tfrac{3}{2}}\label{eq:S squared whittle}
\end{equation}
when $\nu>-3/2$ where $X\equiv kl$.

\begin{figure}
\begin{centering}
\begin{overpic}[width=0.21\columnwidth]{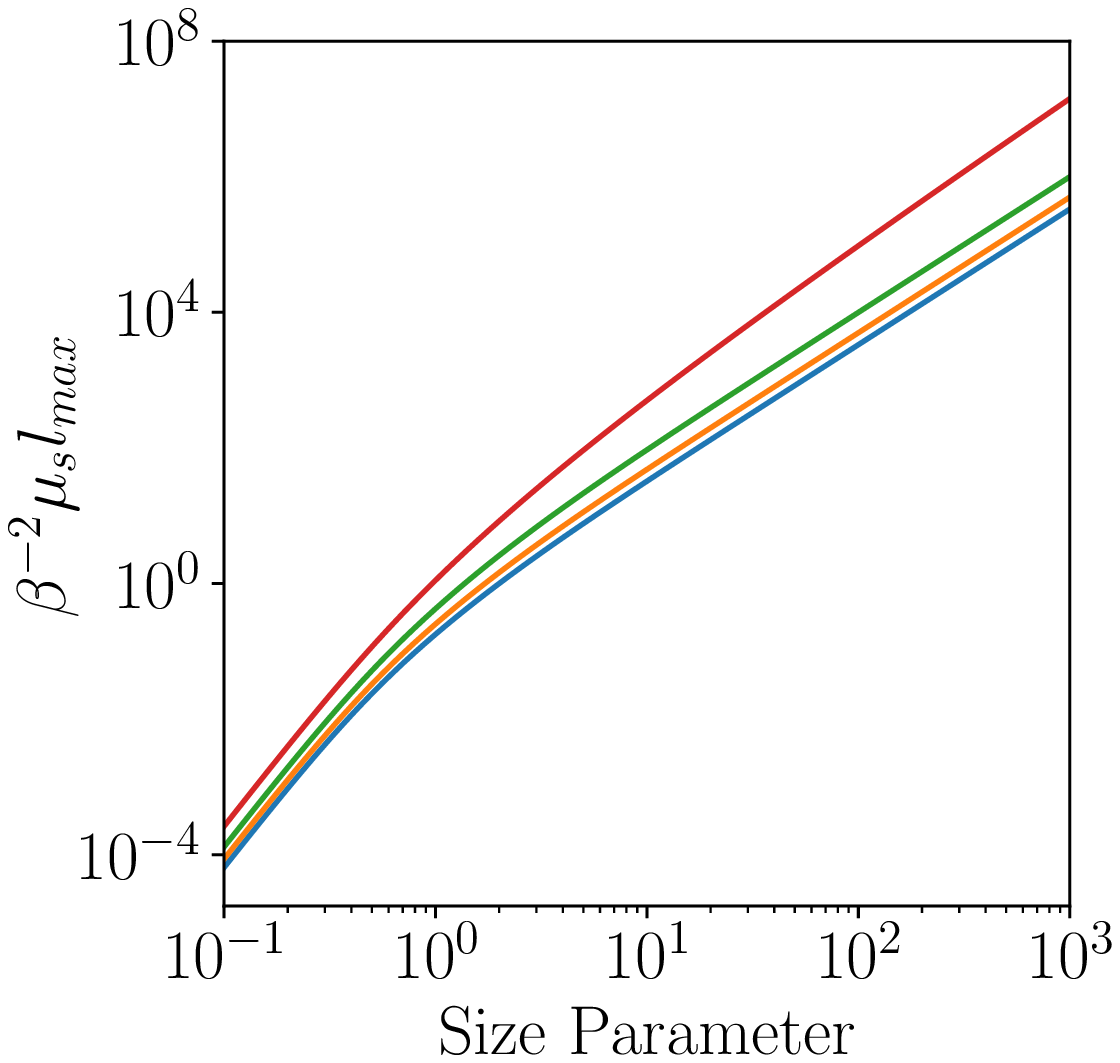}\end{overpic}~\begin{overpic}[width=0.215\columnwidth]{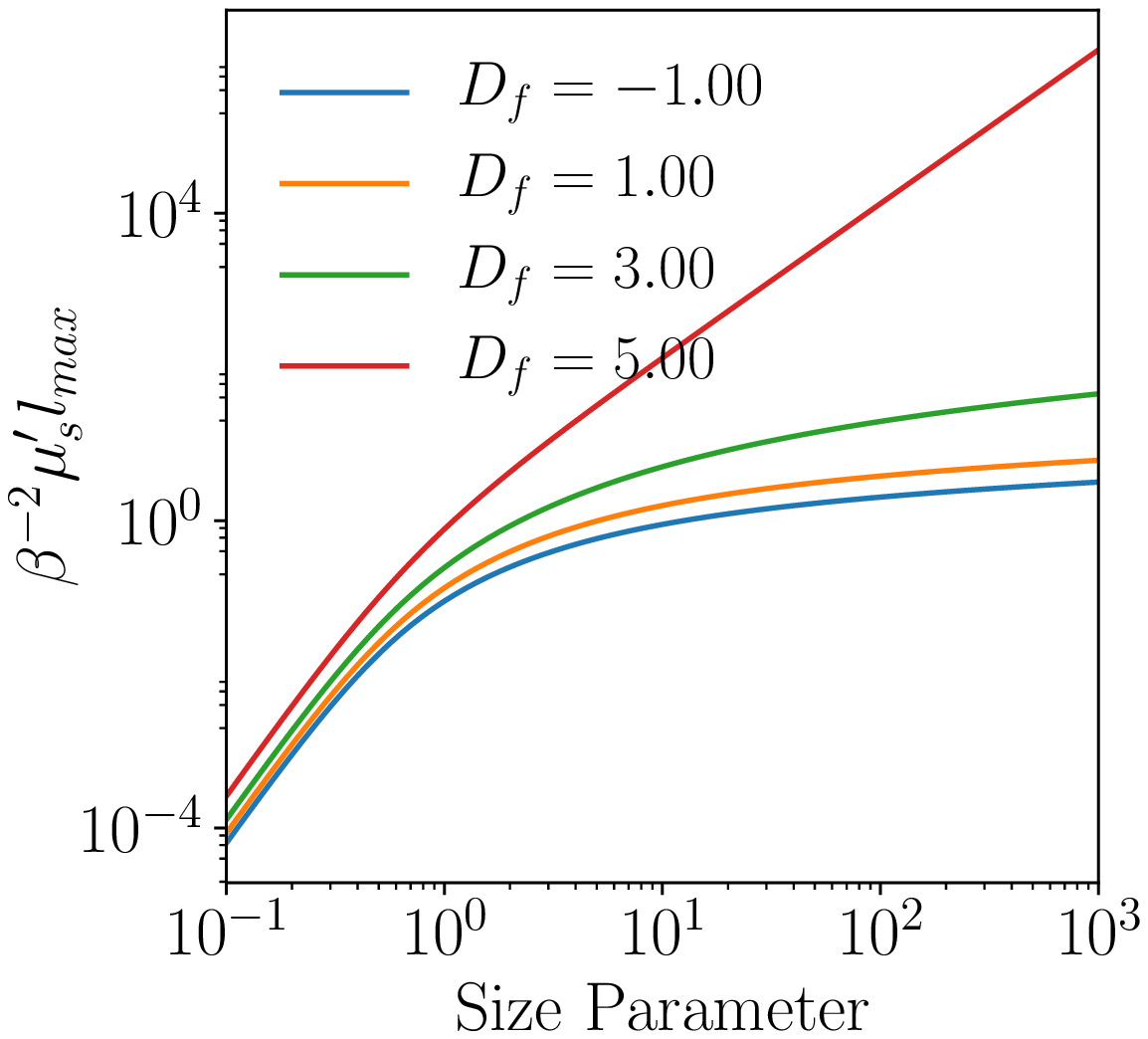}\end{overpic}~\begin{overpic}[width=0.199\columnwidth]{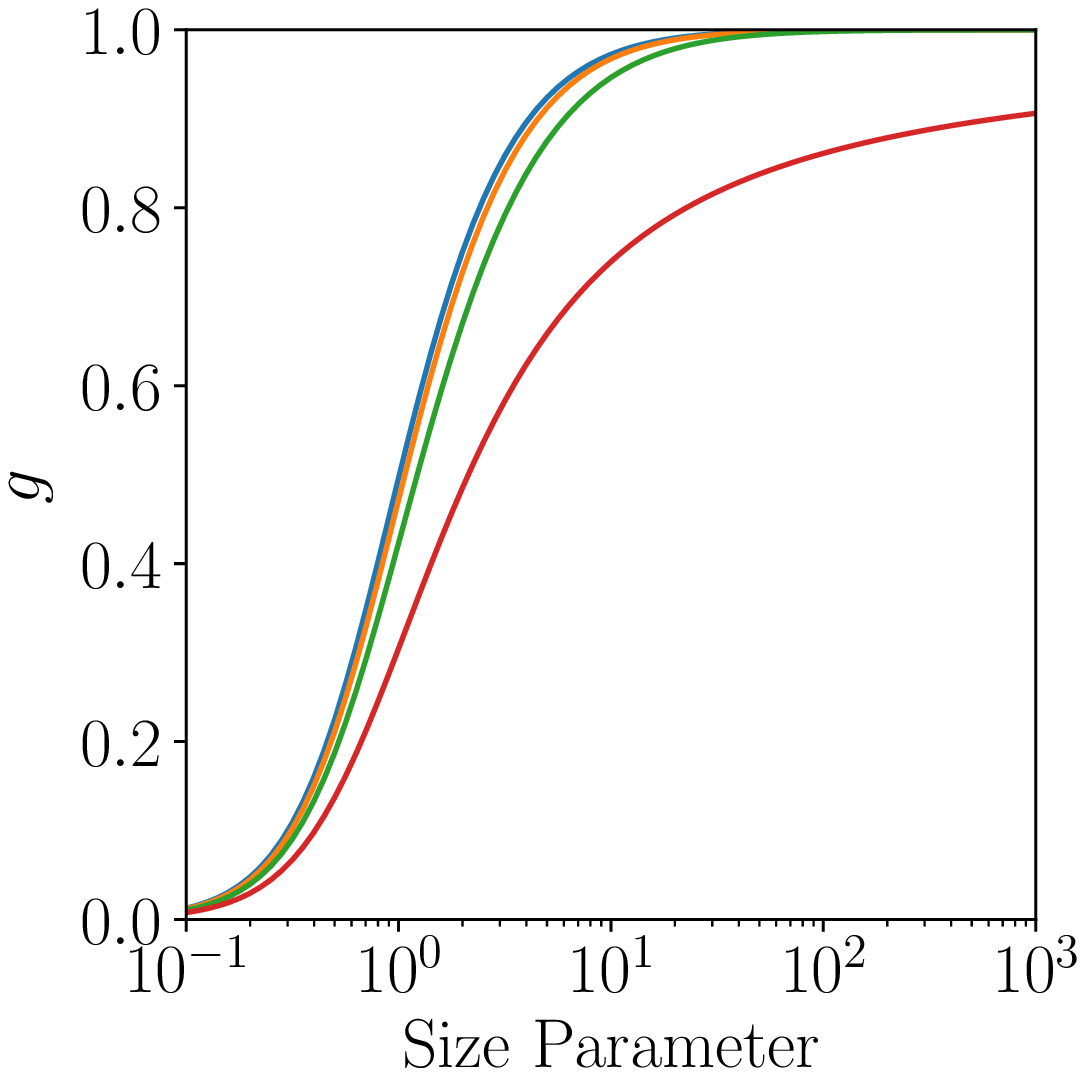}\end{overpic} \begin{overpic}[width=0.276\columnwidth]{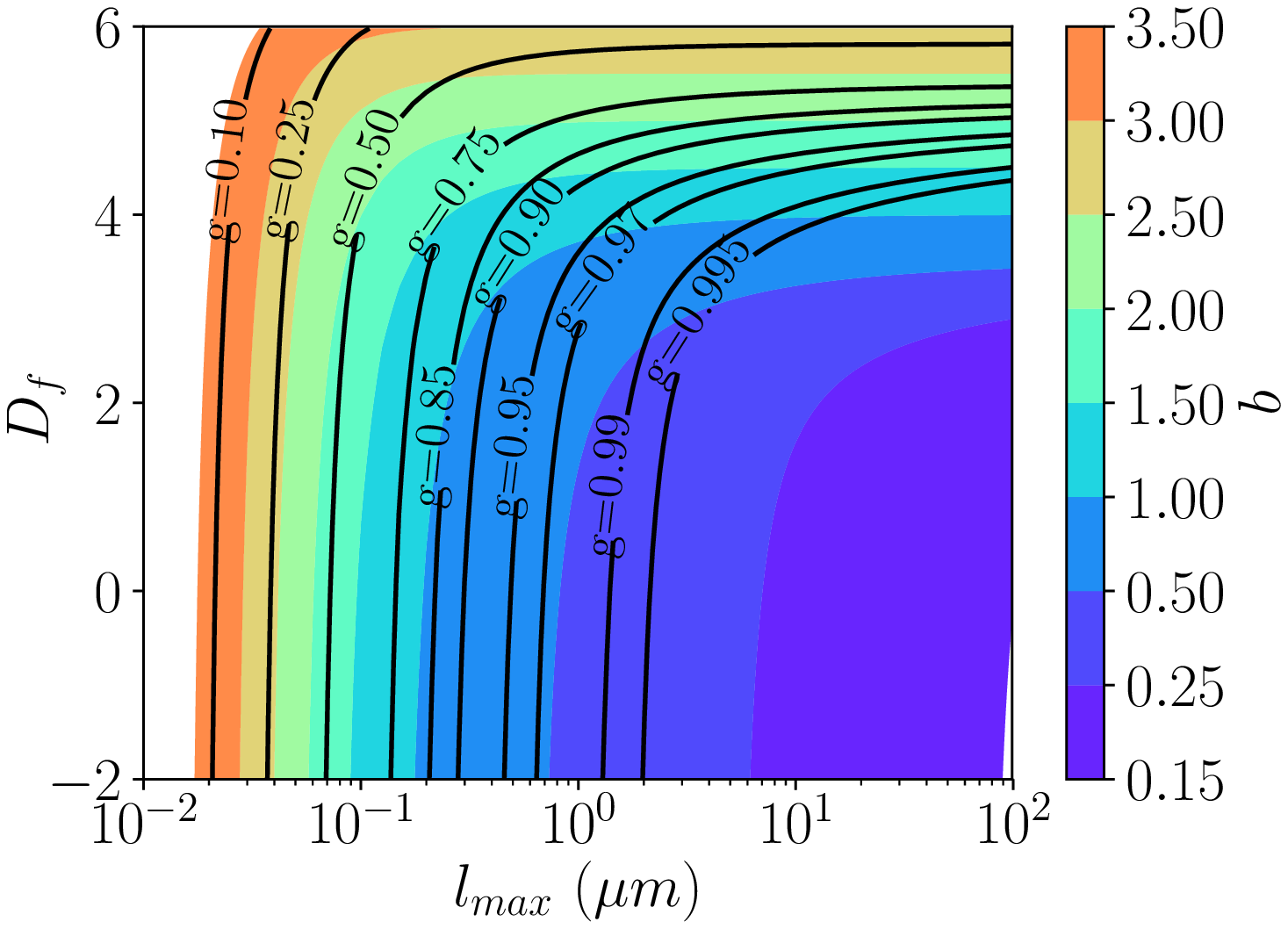}\end{overpic}
\par\end{centering}

\begin{centering}
\begin{overpic}[width=0.21\columnwidth]{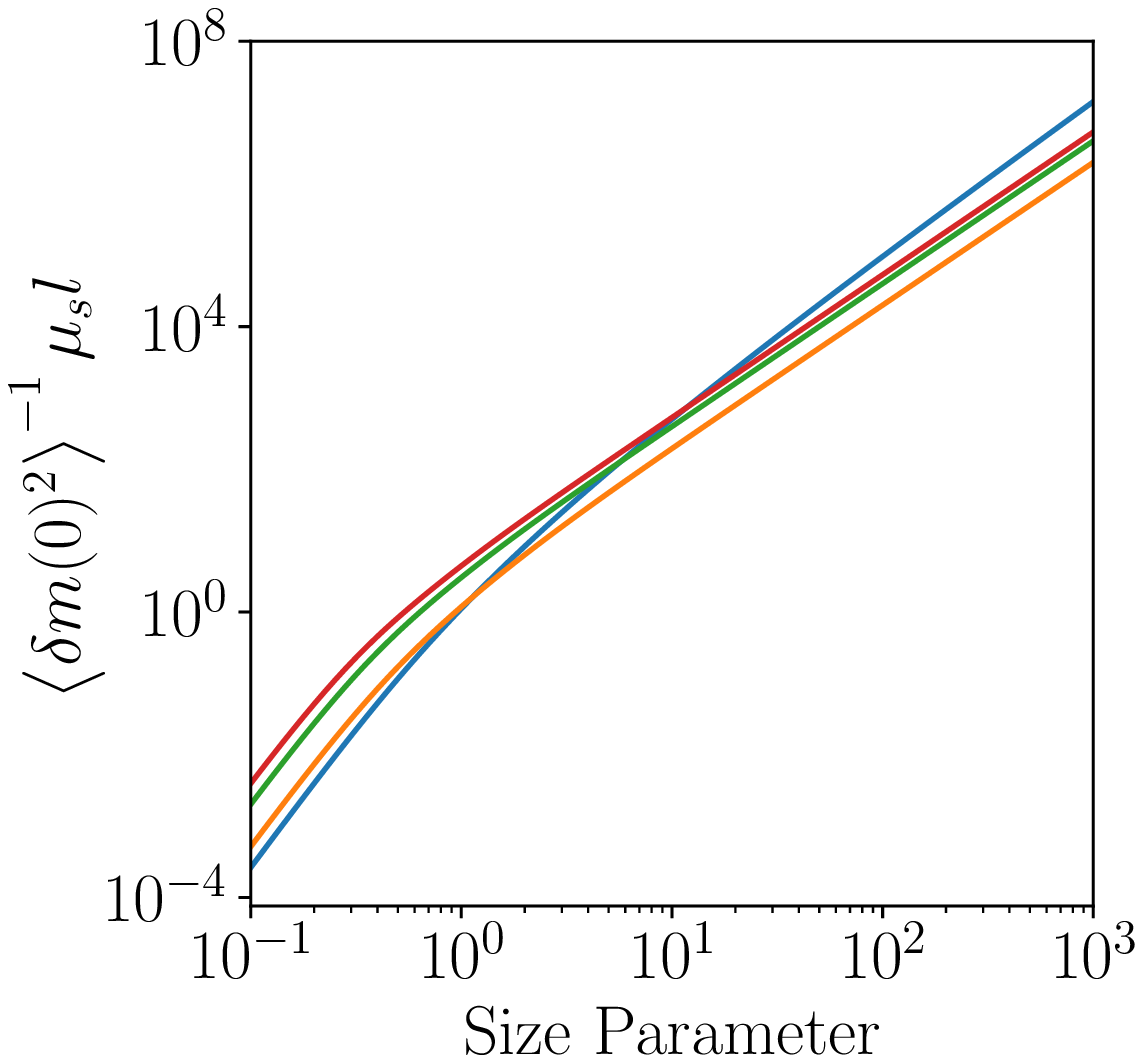}\end{overpic}~\begin{overpic}[width=0.215\columnwidth]{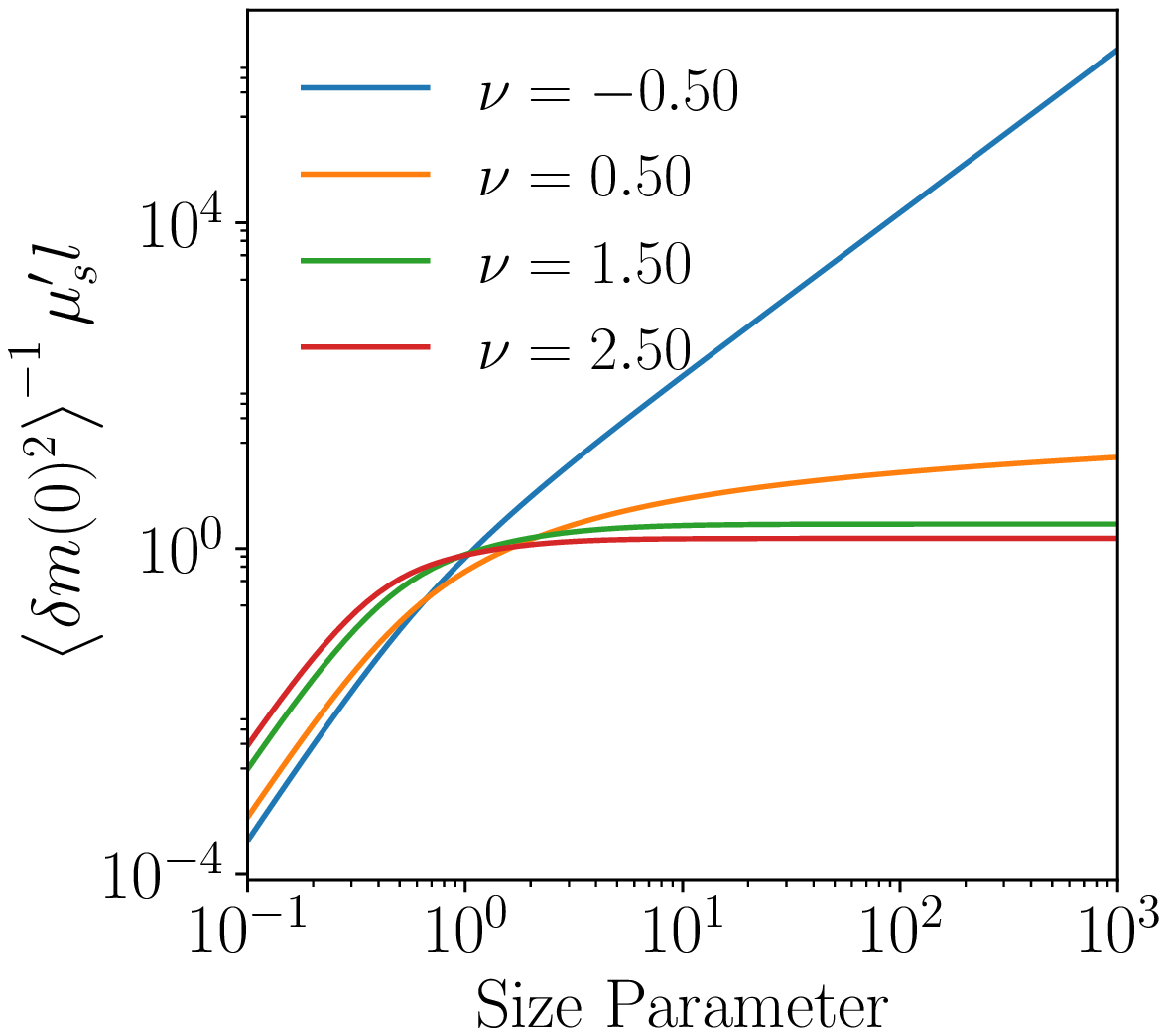}\end{overpic}~\begin{overpic}[width=0.199\columnwidth]{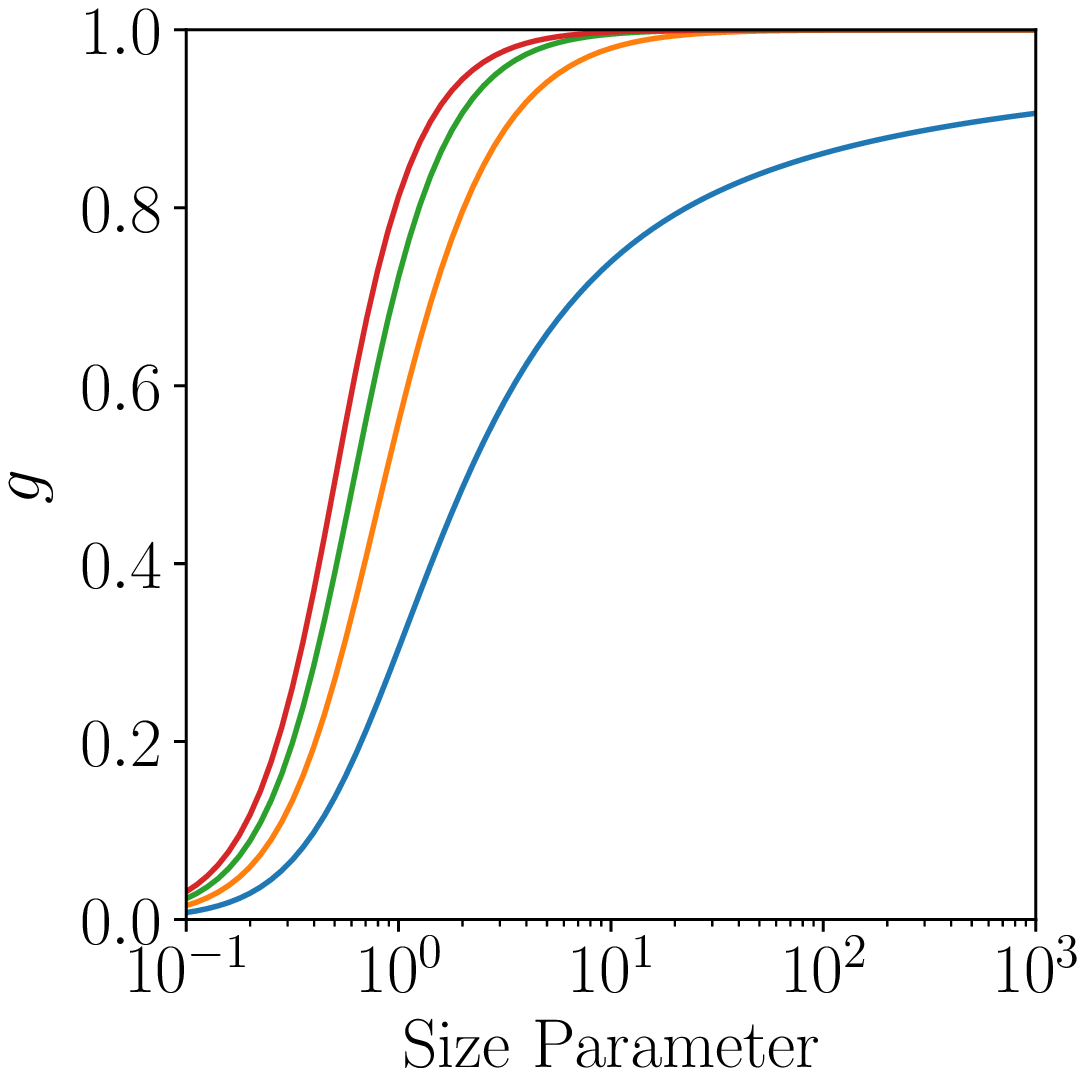}\end{overpic} \begin{overpic}[width=0.276\columnwidth]{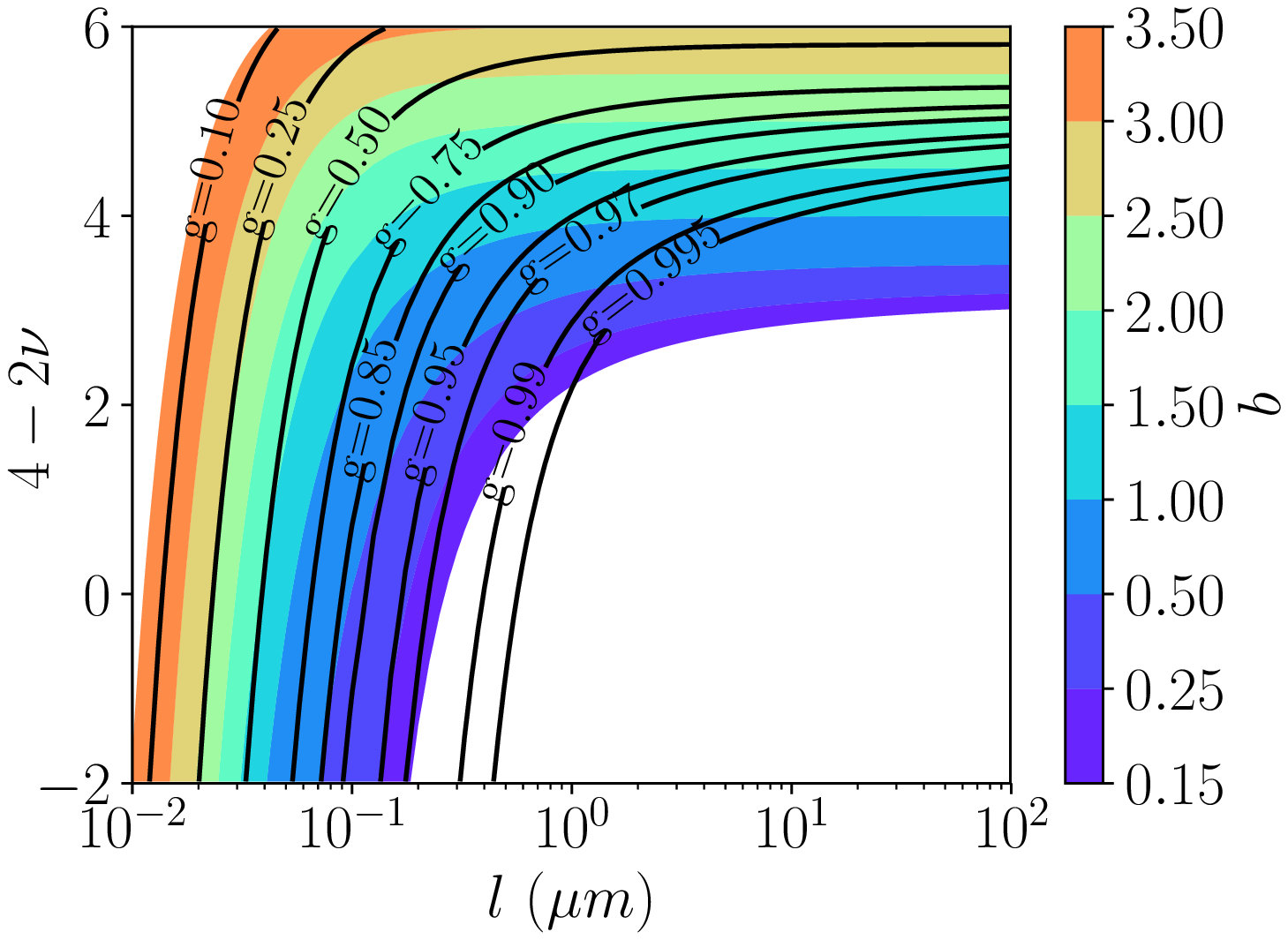}\end{overpic}
\par\end{centering}

\caption{The normalized scattering coefficient $\beta^{-2}\mu_{s}l_{\mathrm{max}}$
($\left\langle \delta m(0)^{2}\right\rangle ^{-1}\mu_{s}l$), reduced
scattering coefficient $\beta^{-2}\mu_{s}'l_{\mathrm{max}}$ ($\left\langle \delta m(0)^{2}\right\rangle ^{-1}\mu_{s}'l$),
the anisotropy factor $g$, and the scattering power $b$ in the fractal
(top row) and Whittle-Matern (bottom row) continuous medium model.
The size parameter is $kl_{\mathrm{max}}$ and $kl$, respectively,
in the two models. The scattering power $b$ is fitted from $\mu_{s}'(\lambda)$
over the spectral window $500\mathrm{nm}<\lambda<700\mathrm{nm}$
centered at $600\mathrm{nm}$. }

\label{fig:fractal vs whittle scattering properties}
\end{figure}

Given the background squared scattering amplitude (\ref{eq:S_bg squared})
and (\ref{eq:S squared whittle}) specified, respectively, in the
fractal and Whittle-Matern models, the scattering properties originating
from the background refractive fluctuation is simply determined by
Eqs.~(\ref{eq:differential csca}-\ref{eq:reduced csca}). Figure
\ref{fig:fractal vs whittle scattering properties} shows the trends
of various scattering properties: the scattering coefficient
$\mu_{s}$, the reduced scattering coefficient $\mu_{s}'$, the
anisotropy factor $g\equiv1-\mu_{s}'/\mu_{s}$,
and the scattering power $b$ ($\mu_{s}'\propto\lambda^{-b}$) predicted
by the two models. The scattering power is computed over the spectral
range from $500\mathrm{nm}$ to $700\mathrm{nm}$ for the center wavelength
$600\mathrm{nm}$.

For all size parameters, the normalized $\mu_{s}$, $\mu_{s}'$, and
$g$ in both fractal and Whittle-Matern models increase with the size
parameter, i.e., $\mu_{s}$, $\mu_{s}'$, and $g$ all decrease with
the wavelength. The dependence of their scattering properties on the
wavelength reduces to an identical power law if interchanging $D_{f}$
and $(4-2\nu)$ when $X\gg1$ (see Appendix \ref{sec:appendix-large-size-limit}). There
are, however, notable differences that first the dependence of the
normalized $\mu_{s}$ and $\mu_{s}'$ on $D_{f}$ in the fractal model
is monotonic across the whole size parameters (increases with $D_{f}$)
whereas it is not the case for the dependence on $\nu$ in the Whittle-Matern
model. Moreover, the scattering power $b$ is more restrictive in
the fractal model than that in the Whittle-Matern model. For example,
$b$ stays above $0.25$ in the fractal model for a medium of $g=0.99$
whereas $b$ can be much smaller than $0.25$ in the Whittle-Matern
model for the medium of an identical anisotropy factor. 

Tissue phantoms which consist of only small particles exhibit similar
trends \cite{Michels2008} as above. Thorough measurements on various
tissue types within visible and near-infrared spectral range \cite{peters.ea90:_optic,Ma2005}
have revealed unexpectedly contradictory trends in, in particular,
$g$, to the above theoretical prediction as well observed by Jacques
in his recent review of optical properties of biological tissues \cite{Jacques2013}.
A pure continuum light scattering model has also been found to be
insufficient in an extensive study of angular light scattering of
water suspensions of human cervical squamous carcinoma epithelial
(HiLa) cells over a wide range of wavelengths (400 to 700nm) \cite{wu07:_unified_mie_and_fract_scatt,xu07:_unified_mie_and_fract_scatt}.

\subsection{Plum pudding random medium}

The deficiencies of the fractal and Whittle-Matern continuous medium
models for tissue light scattering demand a reexamination of the nature
of tissue light scattering. Although the refractive index distribution
in tissue resembles turbulence yet it is \emph{not} a turbulence.
Some prominent structures such as the cell nuclear structure of much
higher refractive index than the background are distinctive from the
surrounding environment. A more realistic picture of tissue is a composite
medium that is a continuum (pudding) yet with some prominent distinctive
structures (plum) embedded inside. The superposition principle of
light scattering by composite particles \cite{xu06:_super} provides
a convenient framework for describing light scattering by such a system.

Light scattering by Plum Pudding Random Medium model of tissue consists
of scattering by distinctive scattering structures and the fluctuation
of the background refractive index. The former includes, for example,
the nuclear structure in soft tissue and fiber bundles in muscle.
The latter incorporates smaller scattering structures such as organelles
and refractive index variations throughout the tissue continuum. The
total squared scattering amplitude, according to the superposition
principle of light scattering by composite particles \cite{xu06:_super,xu07:_unified_mie_and_fract_scatt},
can be written as:
\begin{equation}
\overline{\left|S(\mathbf{q})\right|^{2}}=\left|S_{\mathrm{core}}(\mathbf{q})\right|^{2}+\left|S_{\mathrm{bg}}(\mathbf{q})\right|^{2}\label{eq:S sq avg}
\end{equation}
where $S_{\mathrm{core}}(\mathbf{q})$ represents the scattering amplitude
function of the prominent distinctive scattering ``cores'' and $S_{\mathrm{bg}}(\mathbf{q})$
represents the random fluctuation of the background refractive index
described, for example, by the fractal random medium model (\ref{eq:S_bg squared})
or the Whittle-Matern model (\ref{eq:S squared whittle}). 

Now consider light scattering by the prominent distinctive cores in
tissue.  The cores are of arbitrary shapes and randomly oriented
in tissue. The core could be assumed to have a spherical shape after
averaging over all these individual ones. Further, the cores can be
regarded as optically soft ($\left|m-1\right|\ll1$ where $m\equiv n_{\mathrm{core}}/n_{\mathrm{bg}}$
is the relative refractive index of the core). To account for the
polydispersity of the cores, the radius of the core is assumed to
follow a lognormal distribution $f(a)$ on the radius $a$ (see Appendix
\ref{appendix:logn-size-distr}). A different form of particle size distribution may be
used. Scatterers of different size distributions 
but of the same effective radius and effective variance behave alike
in their properties of light scattering \cite{mishchenko2002:_scattering}.
 The scattering efficiencies for soft particles following
the lognormal size distribution of parameters $\bar{a}$ and $\delta$
are given by:
\begin{equation}
\bar{Q}_{\mathrm{sca}}(\bar{x},m,\delta)=\bar{a}^{-2}\int a^{2}Q_{\mathrm{sca}}(ka,m)f(a)da,
\end{equation}

\begin{equation}
\bar{Q}_{\mathrm{sca}}'(\bar{x},m,\delta)=\bar{a}^{-2}\int a^{2}Q_{\mathrm{sca}}'(ka,m)f(a)da
\end{equation}
respectively where $\bar{x}\equiv k\bar{a}$. Mie theory \cite{wiscombe80:_improv_mie}
can be used to compute these efficiencies in general and empirical
expressions are given in Appendix \ref{appendix:empir-expr-light}. The corresponding
scattering cross section is given by $\pi\bar{a}^{2}\bar{Q}_{\mathrm{sca}}$
etc.

\begin{figure}
\begin{centering}
\includegraphics[width=0.23\columnwidth]{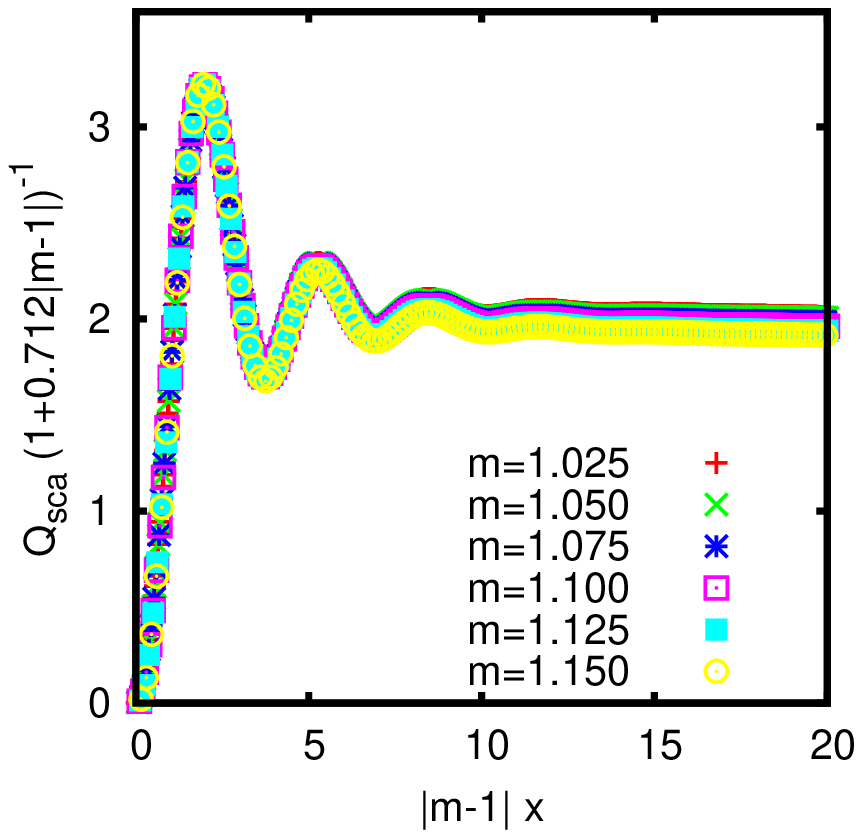}~\includegraphics[width=0.23\columnwidth]{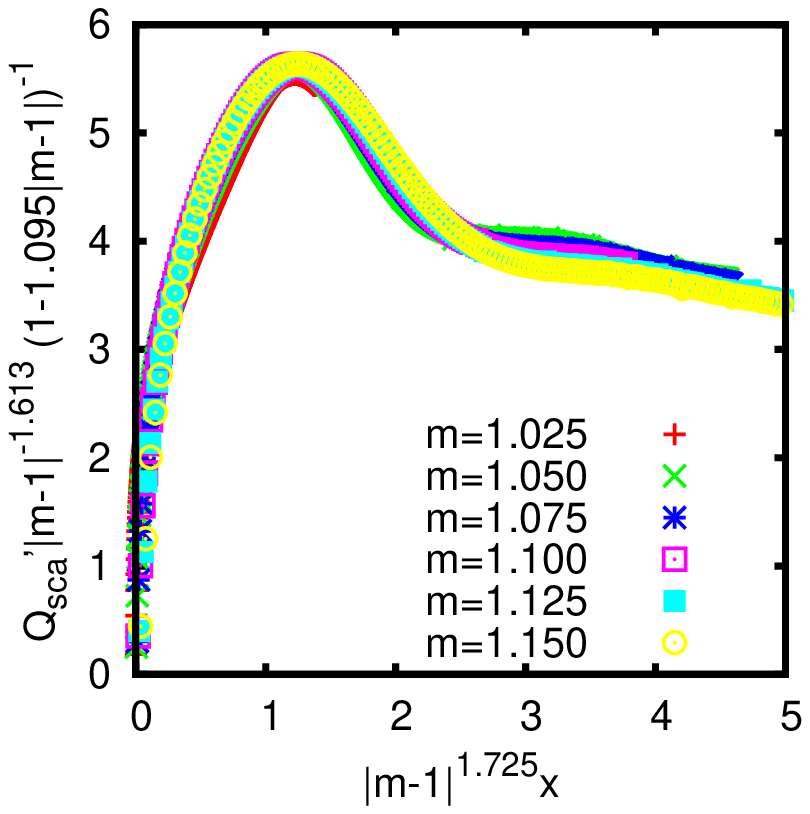}~\includegraphics[width=0.24\columnwidth]{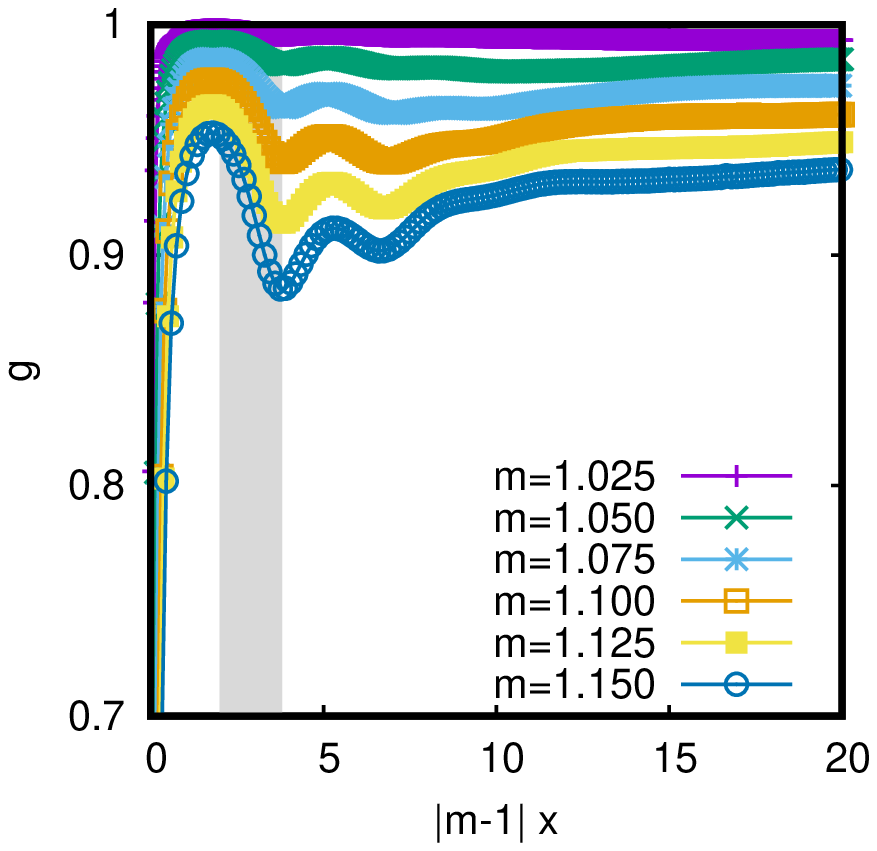}
\par\end{centering}

\caption{The scattering efficiencies collapse approximately to one universal
curve for $\left|m-1\right|\le0.15$ and $0<x<200$ after proper scaling.
The variations in the anisotropy factor increases with the refractive
index $m$. The shaded region in $g$ within $2 \le
\left|m-1\right|^{-1}x \le 3.8$ corresponds to the ``anomalous'' increasing anisotropy factor
with the wavelength within the visible and near-infrared spectral
range observed in most biological tissues.}

\label{fig:polydisperse Qsca etc collapse}
\end{figure}

The polydispersity of the cores tends to smooth and remove the Mie ripples in the spectral dependence
of their scattering properties. To gain insight into the scattering
characteristics of such polydisperse soft particles, one representative
case of the effective size variance $\nu^{\mathrm{eff}}=1.0\%$ ($\delta=0.1$)
is shown in Fig.~\ref{fig:polydisperse Qsca etc collapse}. The scattering
efficiencies collapse approximately to one universal curve respectively
after proper scaling. From this similarity, empirical expressions
have been obtained (see Appendix \ref{appendix:empir-expr-light}). It is clear
that the wavelength dependence of the scattering coefficients and
the anisotropy factor for such soft particles are not monotonic.
There exists multiple regions where their values decrease with the
size parameter and increase with the wavelength. In particular, the
efficiencies $\bar{Q}_{\mathrm{sca}}$ and $\bar{Q}_{\mathrm{sca}}'$
 reach their maximal values at a size parameter of $2\left|m-1\right|^{-1}$
and $1.245\left|m-1\right|^{-1.725}$,  respectively, and the
anisotropy factor $g$ decreases with the size parameter (and increases
with the wavelength) within $2 \le \left|m-1\right|^{-1}x \le 3.8$.

Based on the relationship (\ref{eq:S sq avg}), the bulk scattering
properties of the composite medium are then described by the summation
from the core (plum) and background fluctuation (pudding) components,
\begin{equation}
\mu_{s}=\mu_{s,\mathrm{core}}+\mu_{s,\mathrm{bg}}=N_{c}\pi\bar{a}_{c}^{2}\bar{Q}_{\mathrm{sca}}(\bar{x}_{c},m_{c},\delta)+\mu_{s,\mathrm{bg}},
\end{equation}

\begin{equation}
\mu_{s}'=\mu_{s,\mathrm{core}}'+\mu_{s,\mathrm{bg}}'=N_{c}\pi\bar{a}_{c}^{2}\bar{Q}_{\mathrm{sca}}'(\bar{x}_{c},m_{c},\delta)+\mu_{s,\mathrm{bg}}'
\end{equation}
where $N_{c}$ is the number density of the core with size parameter
$\bar{x}_{c}=k\bar{a}_{c}$, the refractive index $m_{c}$ relative
to the background, and of a lognormal size distribution with parameter
$\bar{a}_{c}$ and $\delta$, and $\mu_{s,\mathrm{bg}}$ and $\mu_{s,\mathrm{bg}}'$
are given in Eqs.~(\ref{eq:csca}, \ref{eq:reduced csca}).  

For most tissues, the size parameter of their core within visible
and near-infrared spectral range is less than $1.245\left|m_{c}-1\right|^{-1.725}$
and resides in the neighborhood of $2\left|m_{c}-1\right|^{-1}$.
The anisotropy factor of light scattering, $g=1-\bar{Q}_{\mathrm{sca}}'/\bar{Q}_{\mathrm{sca}}$,
for the core may increase with the size parameter and decreases with
the wavelength (when $\bar{x}_{c}<2\left|m_{c}-1\right|^{-1}$) or
decrease with the size parameter and increase with the wavelength
(when $2\left|m_{c}-1\right|^{-1} < \bar{x}_{c} < 3.8\left|m_{c}-1\right|^{-1}$). The core in many
tissue types belongs to the latter, responsible for the observed ``anomalous''
anisotropy trend of tissue light scattering (see
Fig.~\ref{fig:polydisperse Qsca etc collapse}). Such cores tend to be 
more dense when the size decreases. The Plum Pudding Random
Medium model for tissue is summarized in Fig.~\ref{fig:plum pudding
  schematic fig}.

\begin{figure}
\begin{centering}
\includegraphics[width=0.3\columnwidth]{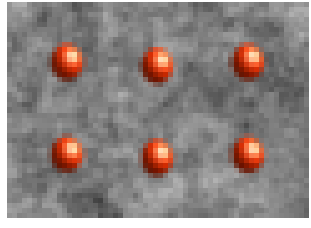}
\par\end{centering}
\caption{The Plum Pudding Random Medium model treats tissue as a composite
medium with some prominent distinctive structures (plum) embedded
inside a continuum (pudding). The former includes, for example, the
nuclear structure in soft tissue. The latter incorporates smaller
scattering structures such as organelles and refractive index variations
throughout the tissue continuum. PPRM faithfully reproduces the wavelength
dependence of tissue light scattering and attributes the \textquotedblleft anomalous\textquotedblright{}
trend in the anisotropy ($g$ increases with the wavelength) to the plum and the powerlaw dependence of
the reduced scattering coefficient on the wavelength to the fractal scattering pudding. }

\label{fig:plum pudding schematic fig}
\end{figure}

It should be noted that the dependence of the scattering efficiency
and the reduced scattering efficiency of the core on the wavelength
is mainly through the product of the power of $(m_{c}-1)$ and the
size parameter, and, in particular, with the former in the form of
$(m_{c}-1)\bar{x}_{c}$ and the latter $(m_{c}-1)^{1.725}\bar{x}_{c}$,
respectively (see Appendix \ref{appendix:empir-expr-light}). The simultaneous knowledge
of scattering and reduced scattering coefficients can thus decouple
$(m_{c}-1)$ and $\bar{x}_{c}$ more reliably than with either parameter
alone, enabling accurate determination of both the refractive index
and the size of the core. 

The determination of the background refractive index fluctuation and
the properties of the core completely characterizes and can further
depict the microstructure and scattering constituents in biological
tissue. That quantification of tissue, although fundamentally from
microscopy of a tissue section, offers the quantitative tissue architecture
and microscopic structure \emph{on average}. We will term that as
\emph{remote microscopy}, derived from non-contact spectroscopic light
scattering measurement on a bulk without tissue excision. 

\section{Results and discussions}

\begin{figure}
\begin{centering}
\includegraphics[width=0.25\textwidth]{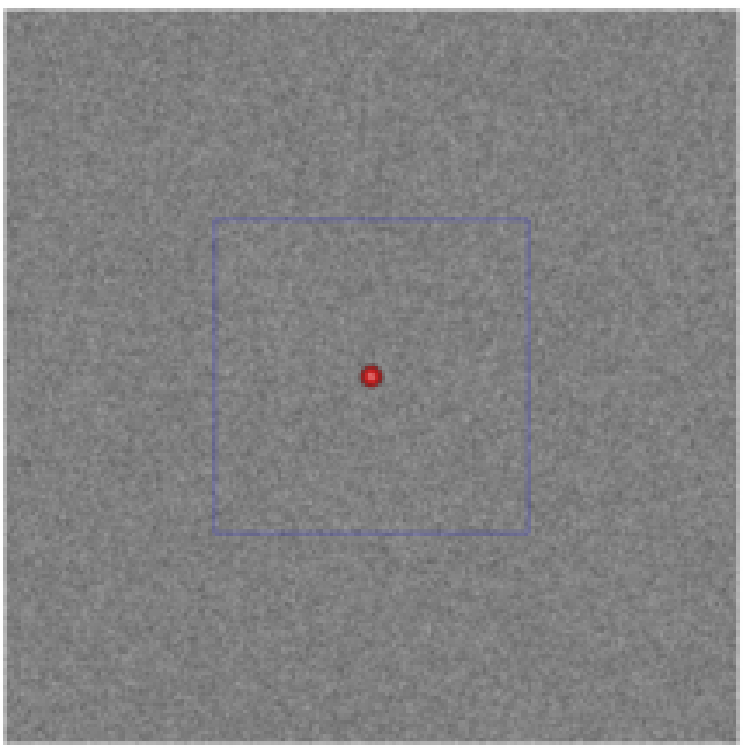}~~\includegraphics[width=0.05125\textwidth]{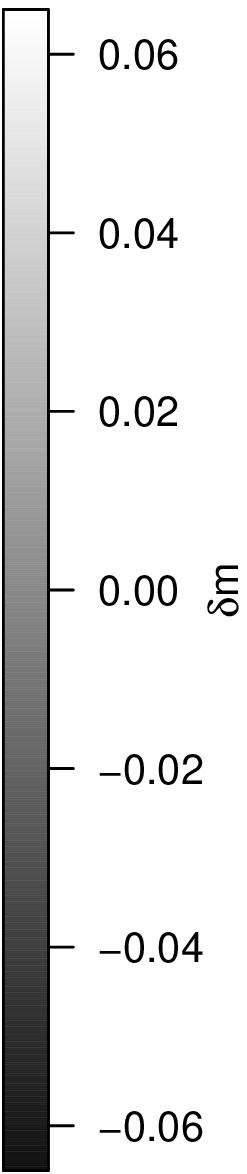}~~
\par\end{centering}

\caption{The plum (core) and pudding (background refractive index fluctuation)
in fresh porcine dermis. The whole window size is $30\mu m\times30\mu m$.
The blue square delineates a unit cell which contains exactly one
core. A core of most probable radius $\bar{a}_{c}\exp(-\delta^{2})$
is shown, surrounded by a shaded area of radius $\bar{a}_{c}\exp(-\delta^{2}+\sqrt{2\log2}\delta)$
at which the number density of the core drops to half maximum. The
core has a relative refractive index $m_{c}=1.172$ (``red'' color). }

\label{fig:core and porcine refractive index fluctuation}
\end{figure}

Figure \ref{fig:porcine-dermis-fitting} shows the PPRM tissue model
fitting \cite{endnote1} (see Appendix \ref{appendix:fitting-procedure}) to the scattering parameters of the fresh porcine dermis tissue
measured from diffuse reflectance and transmission using an integrating
sphere \cite{Ma2005,BASHKATOV2011}. Neither the plum nor the pudding
alone can fit the data. The parameters from the PPRM fitting 
is summarized in Table \ref{tab:porcine parameters}. The fitted parameters
fully characterizes the tissue and provides microscopic details on
the underlying scattering structure. The background refractive index
fluctuation has a maximum correlation length $l_{\mathrm{max}}=0.308\mu m$
and fractal dimension $D_{f}=6.58$. The effective amplitude of the
fluctuation is $\beta=0.545\times10^{-3}$, which depends on both
the amplitude $\sqrt{\left\langle \delta m(0)^{2}\right\rangle }$
and the inner scale $l_{\mathrm{min}}$ of the background refractive
index fluctuation. The value of $\beta$ yields $\sqrt{\left\langle \delta m(0)^{2}\right\rangle }=0.0115$
assuming the inner scale $l_{\mathrm{min}}=20\mathrm{nm}$ for the
background refractive index fluctuation. This inner scale corresponds
to the size of the smallest structure in tissue \cite{schmittkumar98:_optical}.
The core in the dermis has a concentration of $N_{c}=0.473\times10^{-3}\mu m^{-3}$,
i.e., one core per cube of size $12.8\mu m$ on average. The core
has an average radius $\bar{a}_{c}=0.915\mu m$ and the relative refractive
to the background $m_{c}=1.172$. The porcine dermis has a refractive
index $n_{\mathrm{bg}}\simeq1.36$ within the spectral range \cite{Ma2005}
and hence $n_{\mathrm{core}}\simeq1.59$. The core may correspond to the nucleolus which has a
substantially higher refractive index than the rest of the
nucleus and is of similar size \cite{Shaw2005,Foster2012}. Another possibility is due to
melanosomes whose refractive index ranges between $1.55-1.65$
\cite{Nielsen2006} as the fresh porcine dermis in the reported
experiment contained melanin \cite{Ma2005}. The plum 
(cores) and pudding (background refractive index fluctuation $\delta m$)
in porcine dermis with parameters specified in Table \ref{tab:porcine parameters}
is shown in Fig.~\ref{fig:core and porcine refractive index fluctuation}.

\begin{figure}
\begin{centering}
\includegraphics[width=0.25\columnwidth]{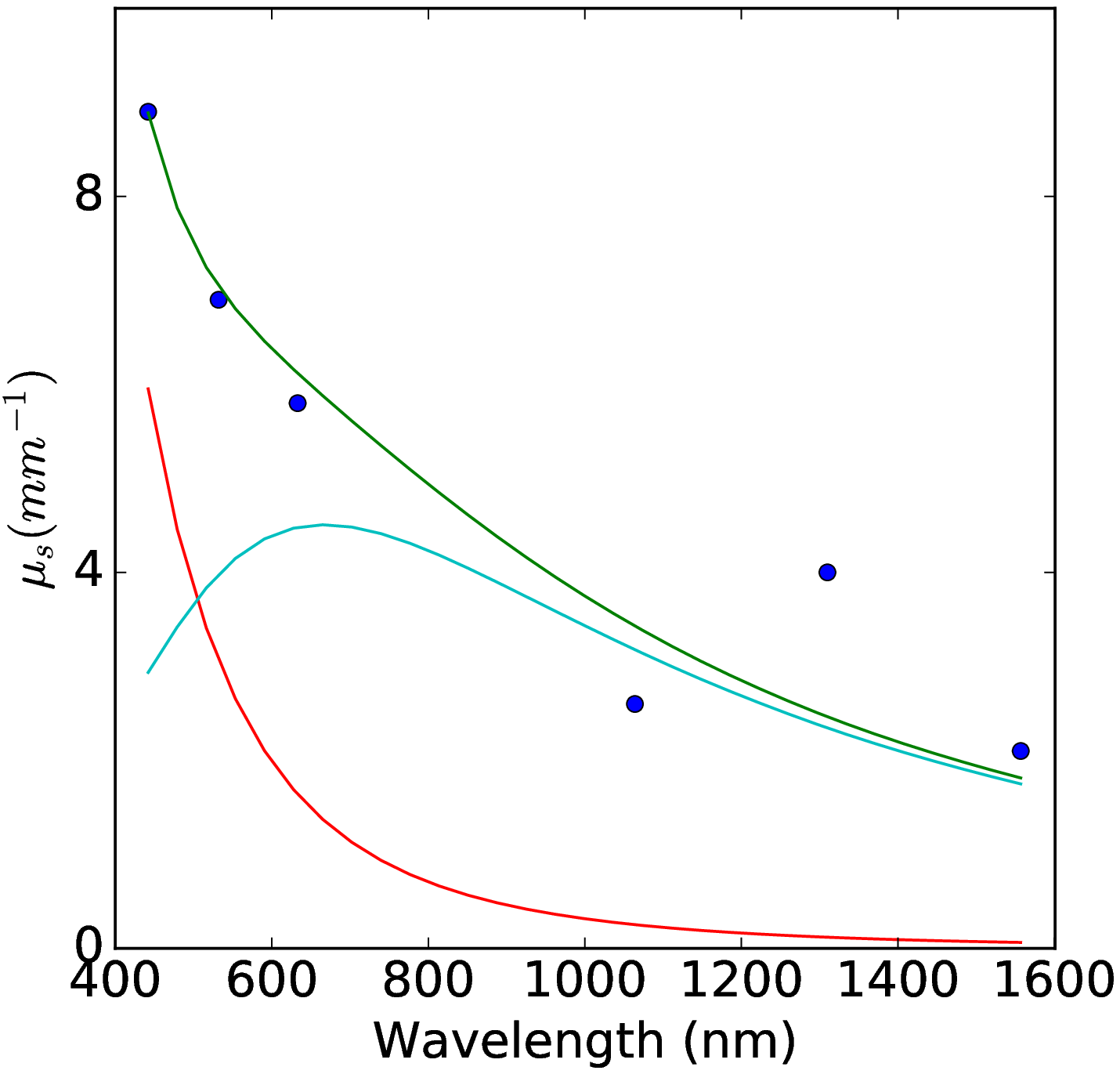}~~\includegraphics[width=0.25\columnwidth]{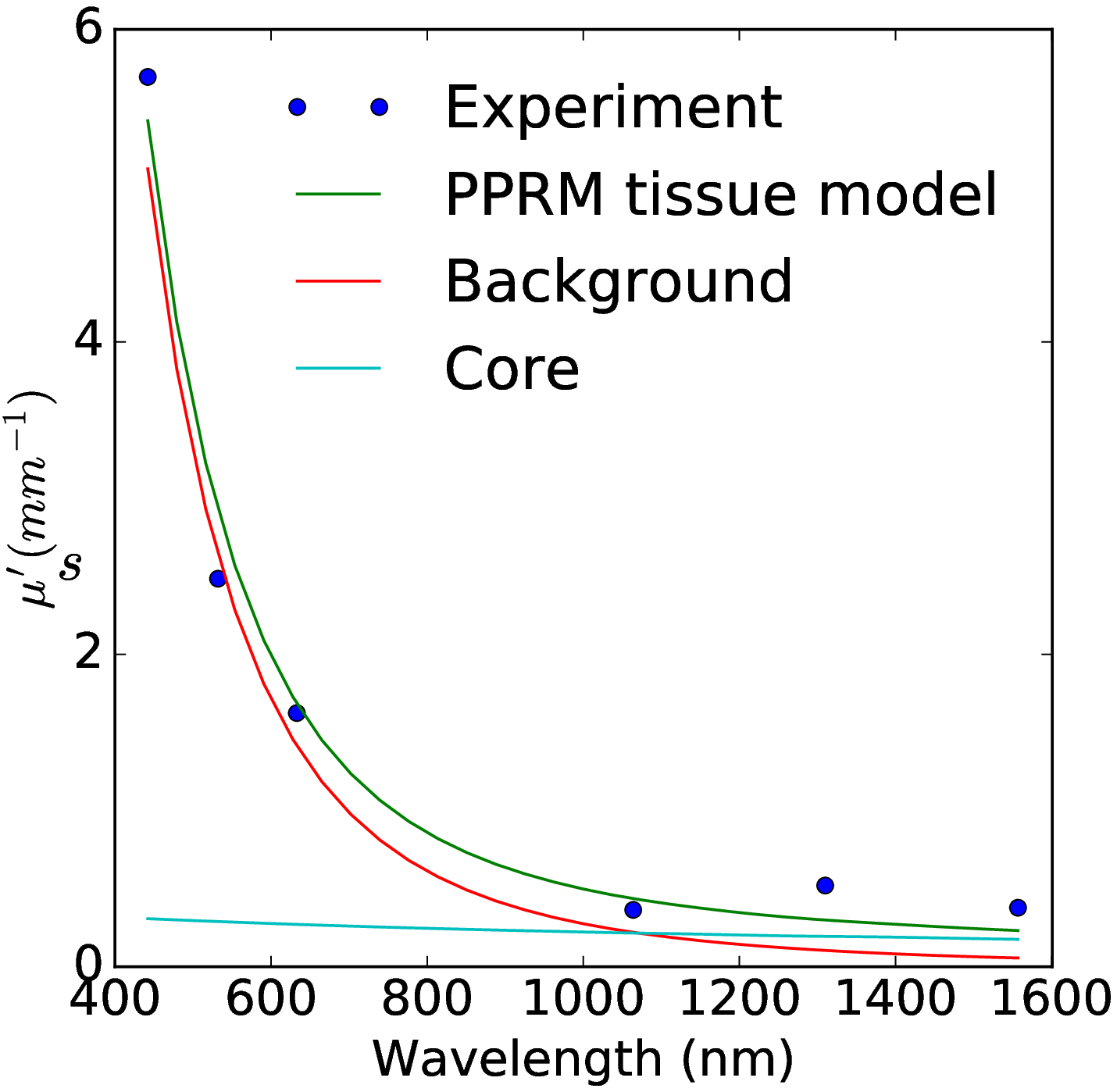}~~\includegraphics[width=0.256\columnwidth]{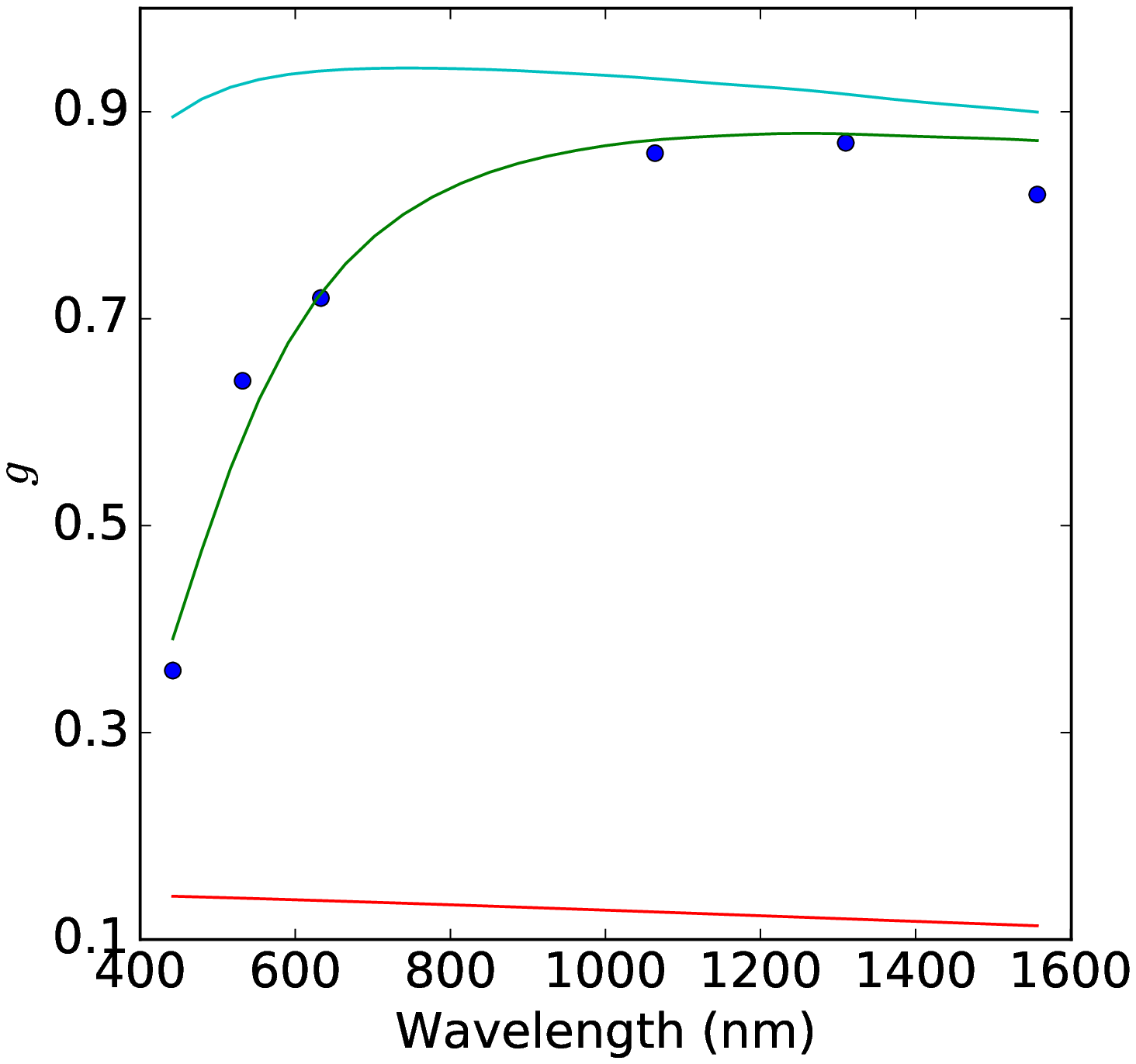}
\par\end{centering}

\caption{Plum pudding random medium tissue model fitting of the fresh porcine
dermis tissue. The columns from left to right show $\mu_{s}$, $\mu_{s}'$
and $g$. The background refractive index fluctuation and the core
are shown together with the PPRM tissue model. Experimental data is
adapted from Ma et al \cite{Ma2005}.}

\label{fig:porcine-dermis-fitting}
\end{figure}

\begin{table*}
\caption{Fitted parameters for fresh porcine dermis tissue. The fluctuation
amplitude $\sqrt{\left\langle \delta m(0)^{2}\right\rangle }$ is
computed from $\beta$ by assuming the inner cutoff for the background
refractive index fluctuations to be $l_{\mathrm{min}}=20\mathrm{nm}$. }

\begin{centering}
\begin{tabular}{ccccccccc}
\multicolumn{3}{c}{Background} & \multicolumn{3}{c}{Core} &  &  & \tabularnewline
$\beta(\times10^{-3})$ & $l_{\mathrm{max}}(\mu m)$ & $D_{f}$ & $N_{c}(\mu m^{-3})$ & $\bar{a}_{c}(\mu m)$ & $m_{c}$ & $\delta$ & $\sqrt{\left\langle \delta m(0)^{2}\right\rangle }$ & error\tabularnewline
\hline 
$0.545$ & $0.308$ & $6.58$ & $0.473\times10^{-3}$ & $0.915$ & $1.172$ & $0.051$ & $0.0115$ & $0.217$\tabularnewline
\end{tabular}
\par\end{centering}

\label{tab:porcine parameters}
\end{table*}

The relative importance of the two components (background vs core,
or, pudding vs plum) in their contributions to $\mu_{s}$ and $\mu_{s}'$
vary significantly with wavelength. For example, the core contributes
$35\%$ to $\mu_{s}$ and the background contributes $94\%$ to $\mu_{s}'$
at $450nm$ whereas the core contributes $96\%$ to $\mu_{s}$ and
the background contributes $30\%$ to $\mu_{s}'$ at $1400nm$. 

We have also attempted to use the Whittle-Matern model for the background
refractive index fluctuation in the PPRM tissue model to fit the experimental
data and were unable to get a satisfactory fitting result. This suggests
although both the fractal and Whittle-Matern model have similar behavior
at the large size parameter limit they behave quite differently at
the low and median size parameter regime and that the fractal model
may describe the tissue background refractive index fluctuation more
accurately. We hence report only the performance of the PPRM model
with the fractal continuum model hereafter. 

\begin{figure}
\begin{centering}
\begin{overpic}[width=0.25\columnwidth]{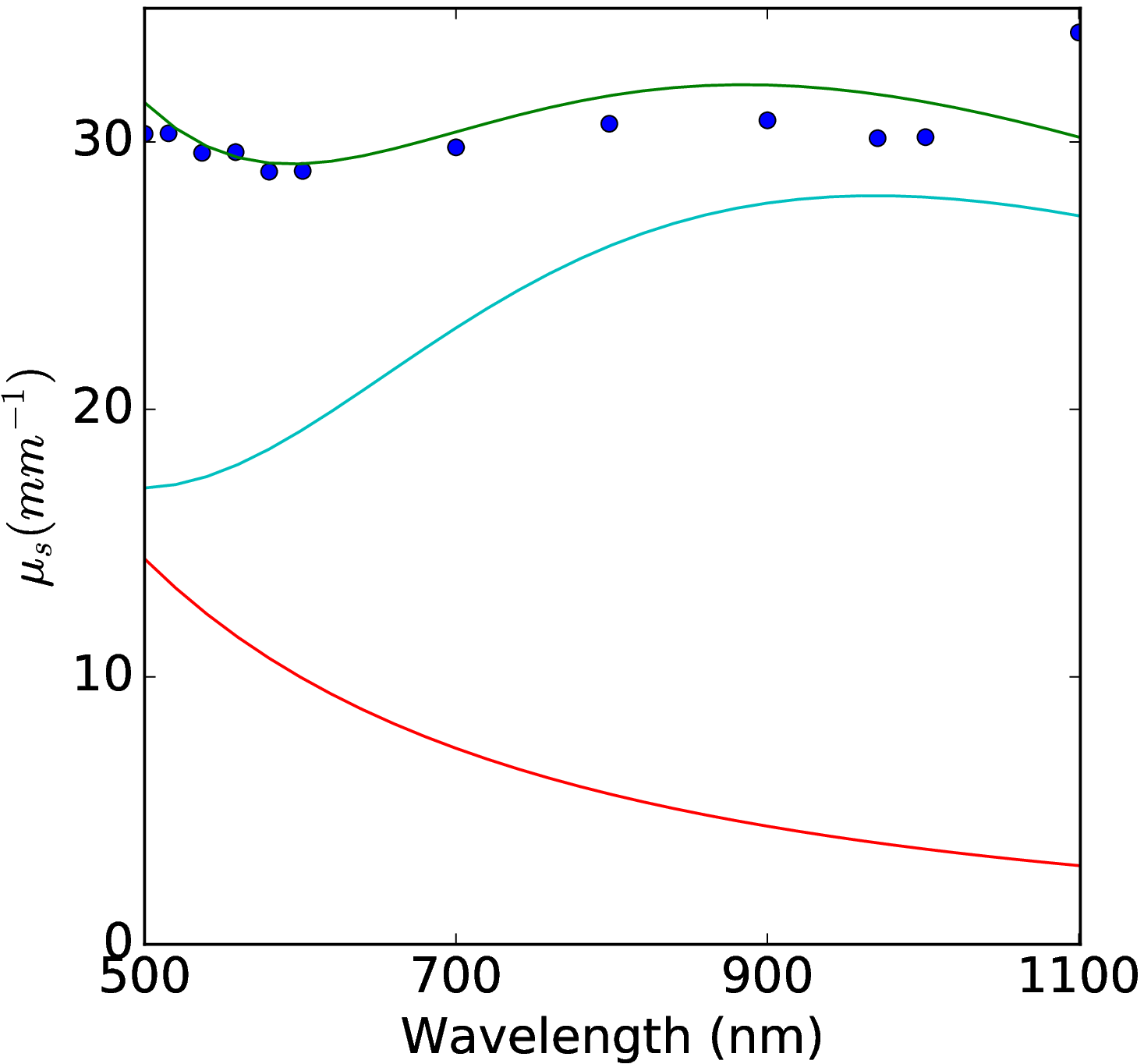}\put(-10,17){\small{(a)}}\end{overpic}~\begin{overpic}[width=0.25\columnwidth]{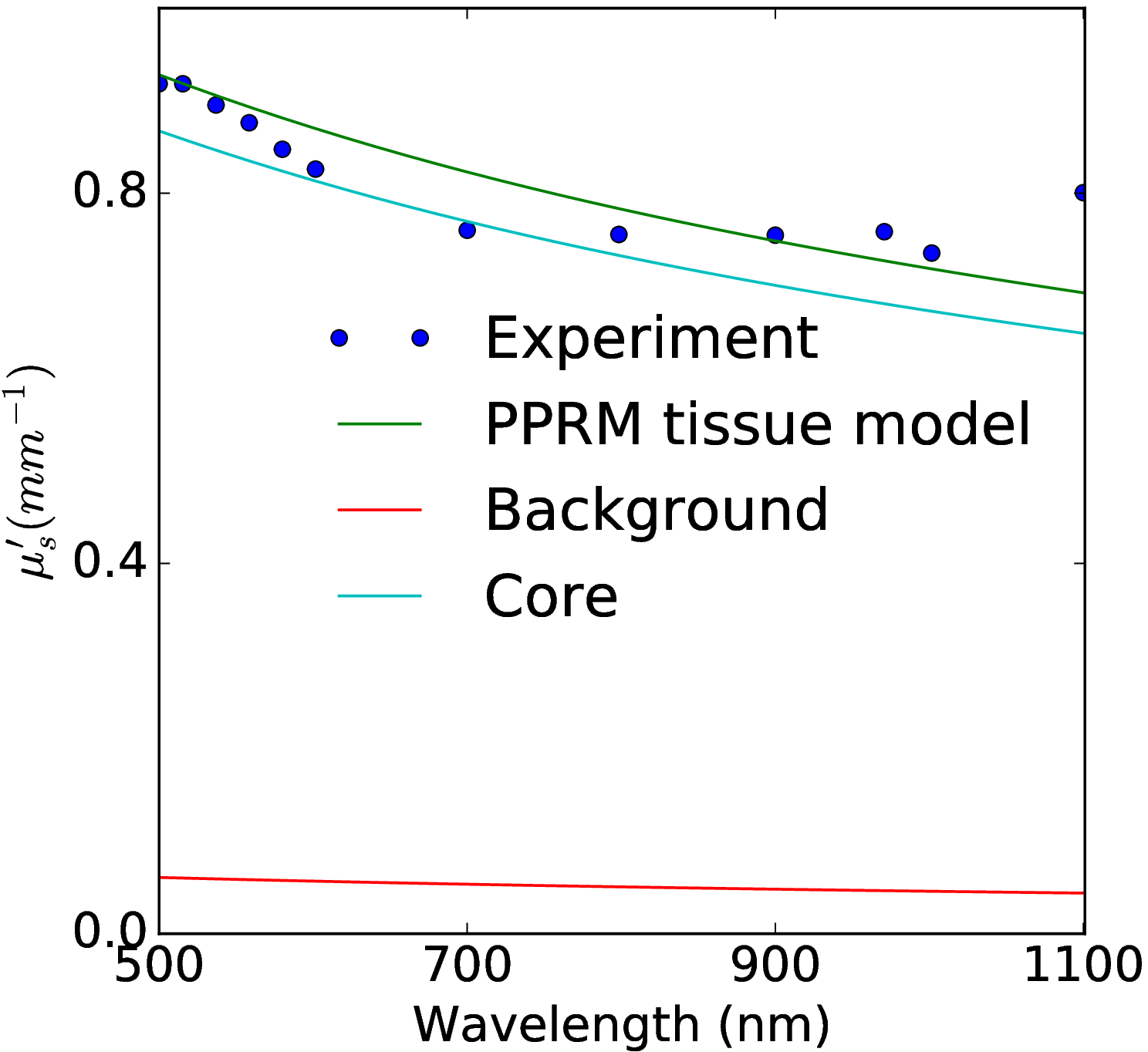}\end{overpic}~\begin{overpic}[width=0.25\columnwidth]{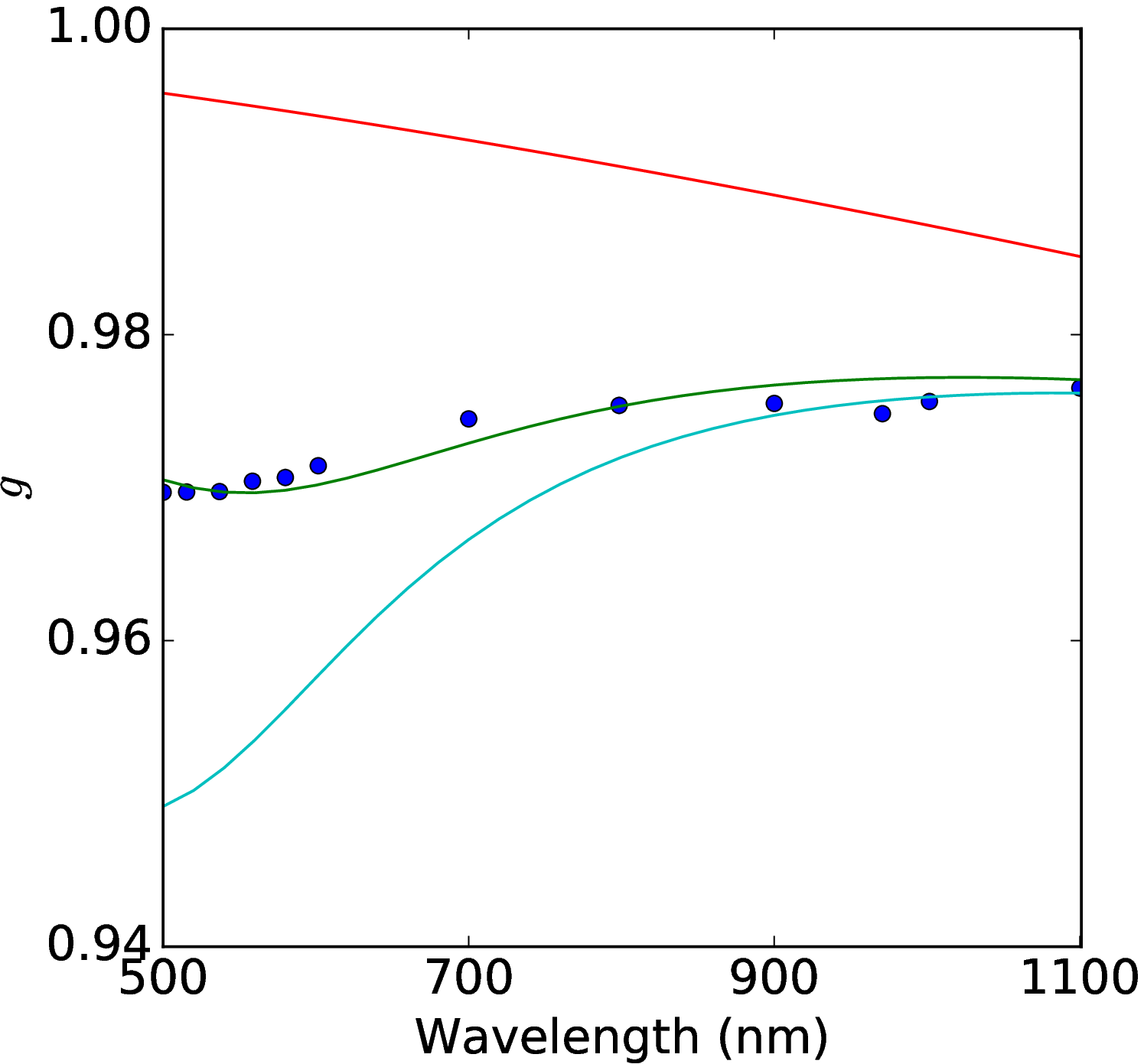}\end{overpic}
\par\end{centering}

\begin{centering}
\begin{overpic}[width=0.25\columnwidth]{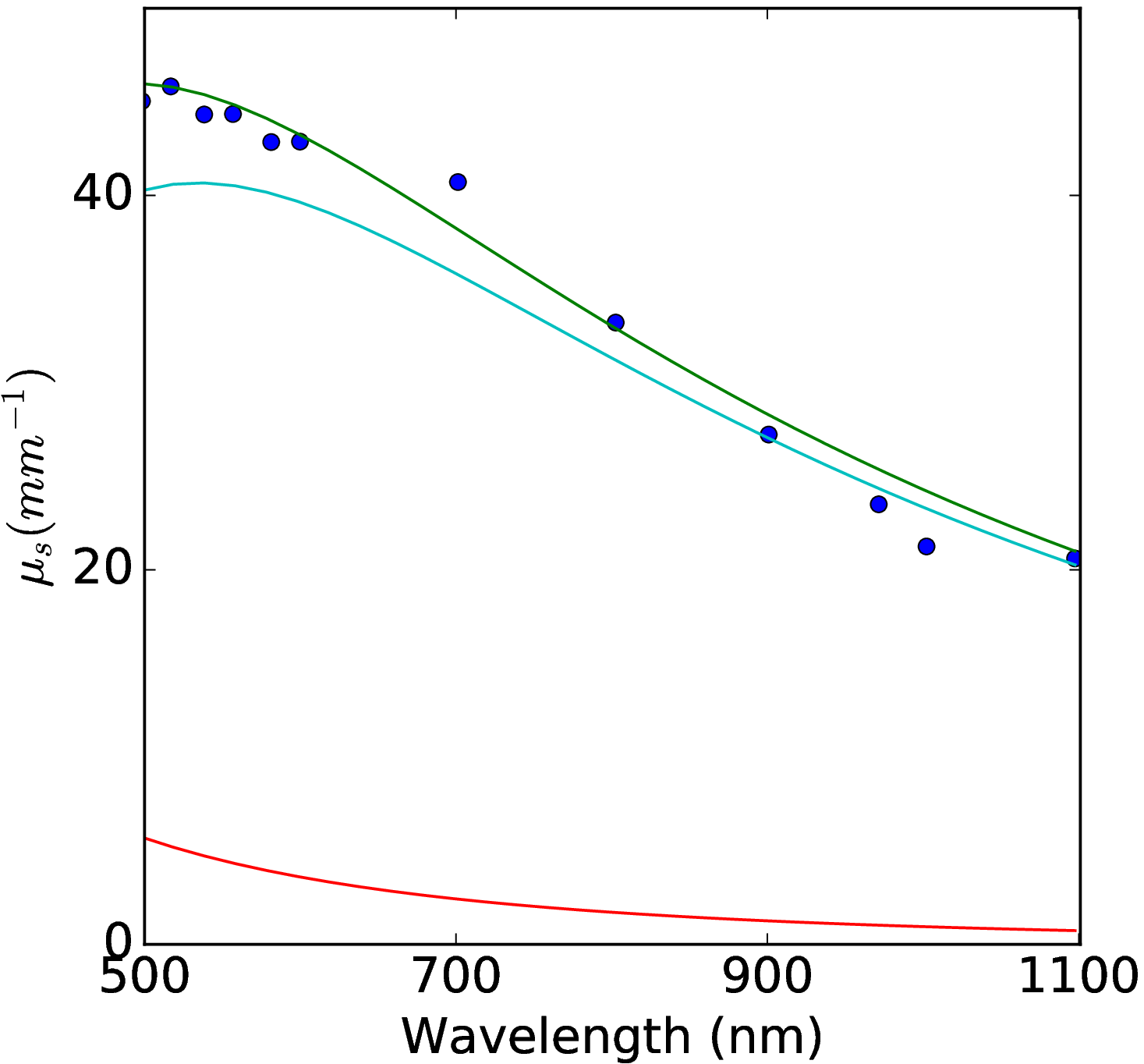}\put(-10,17){\small{(b)}}\end{overpic}~\begin{overpic}[width=0.25\columnwidth]{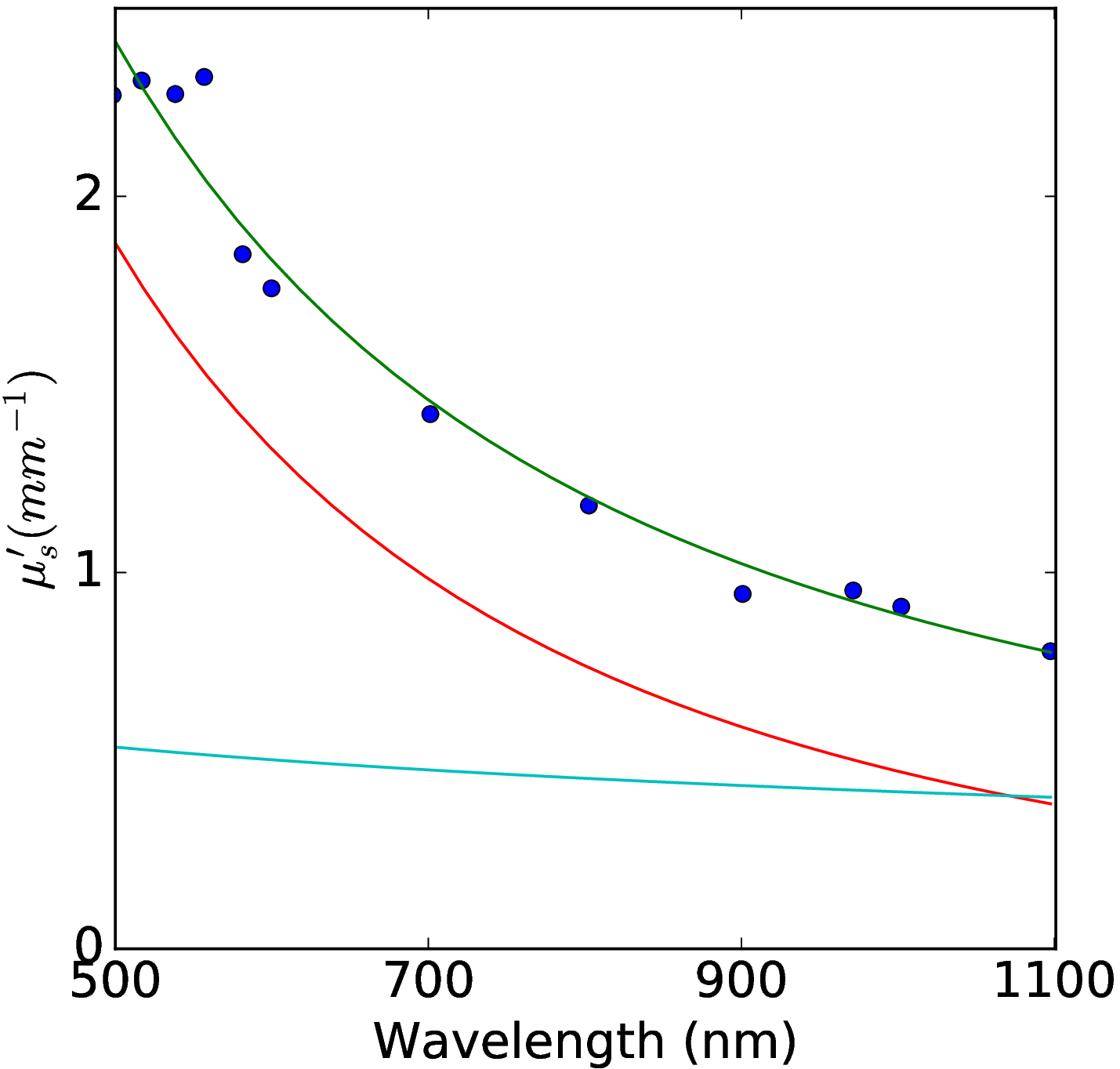}\end{overpic}~\begin{overpic}[width=0.25\columnwidth]{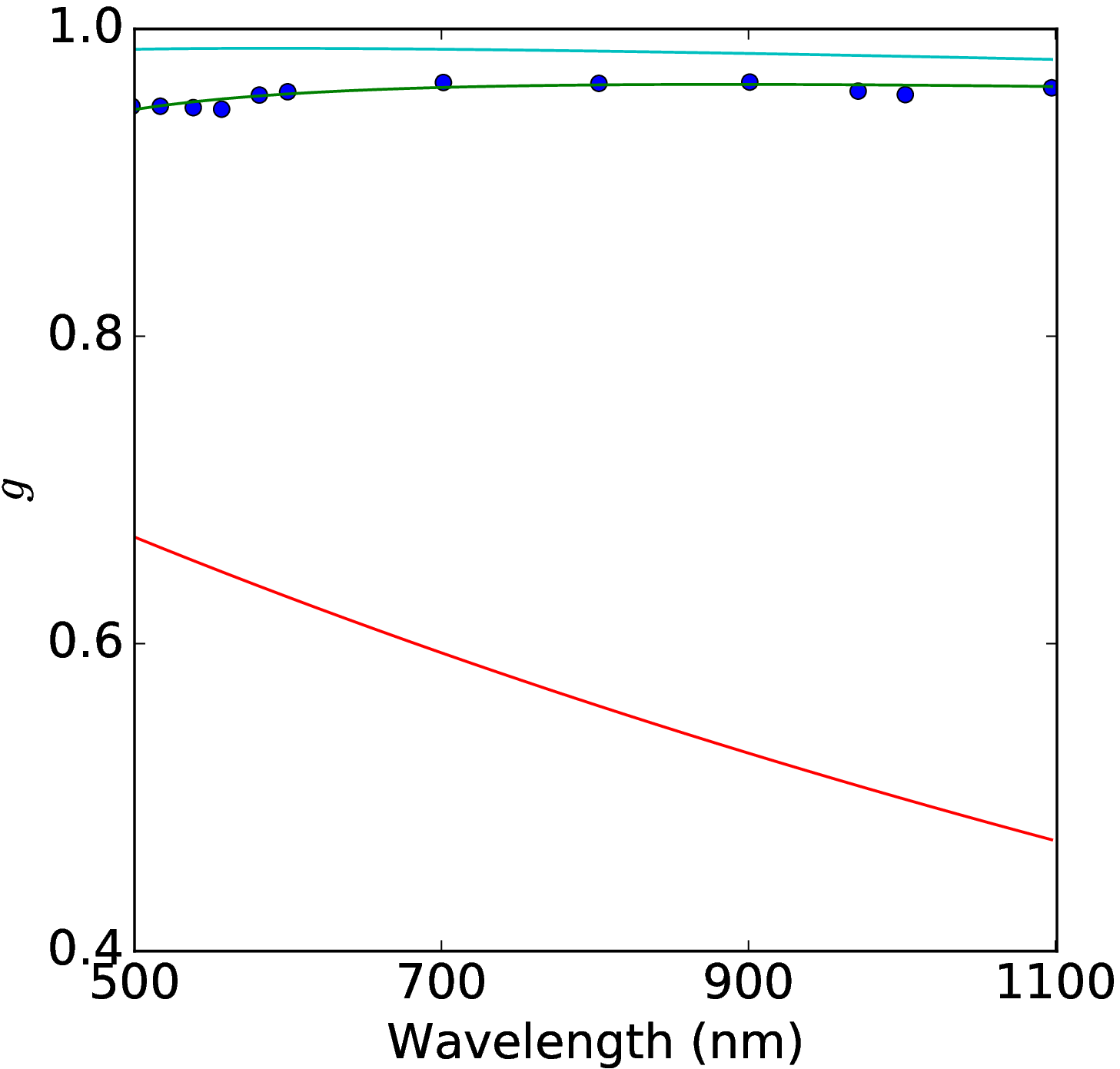}\end{overpic}
\par\end{centering}

\begin{centering}
\begin{overpic}[width=0.25\columnwidth]{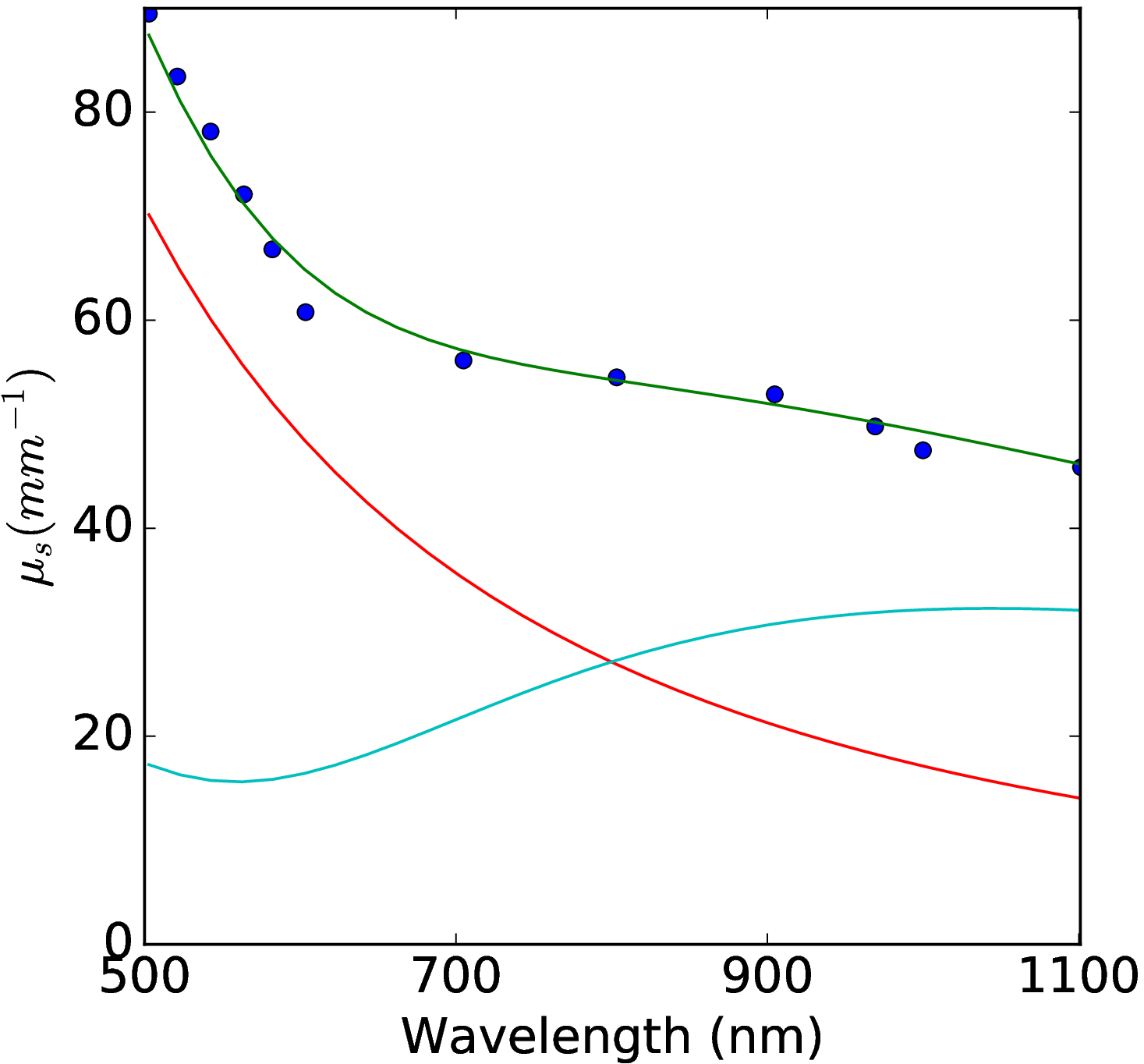}\put(-10,17){\small{(c)}}\end{overpic}~\begin{overpic}[width=0.25\columnwidth]{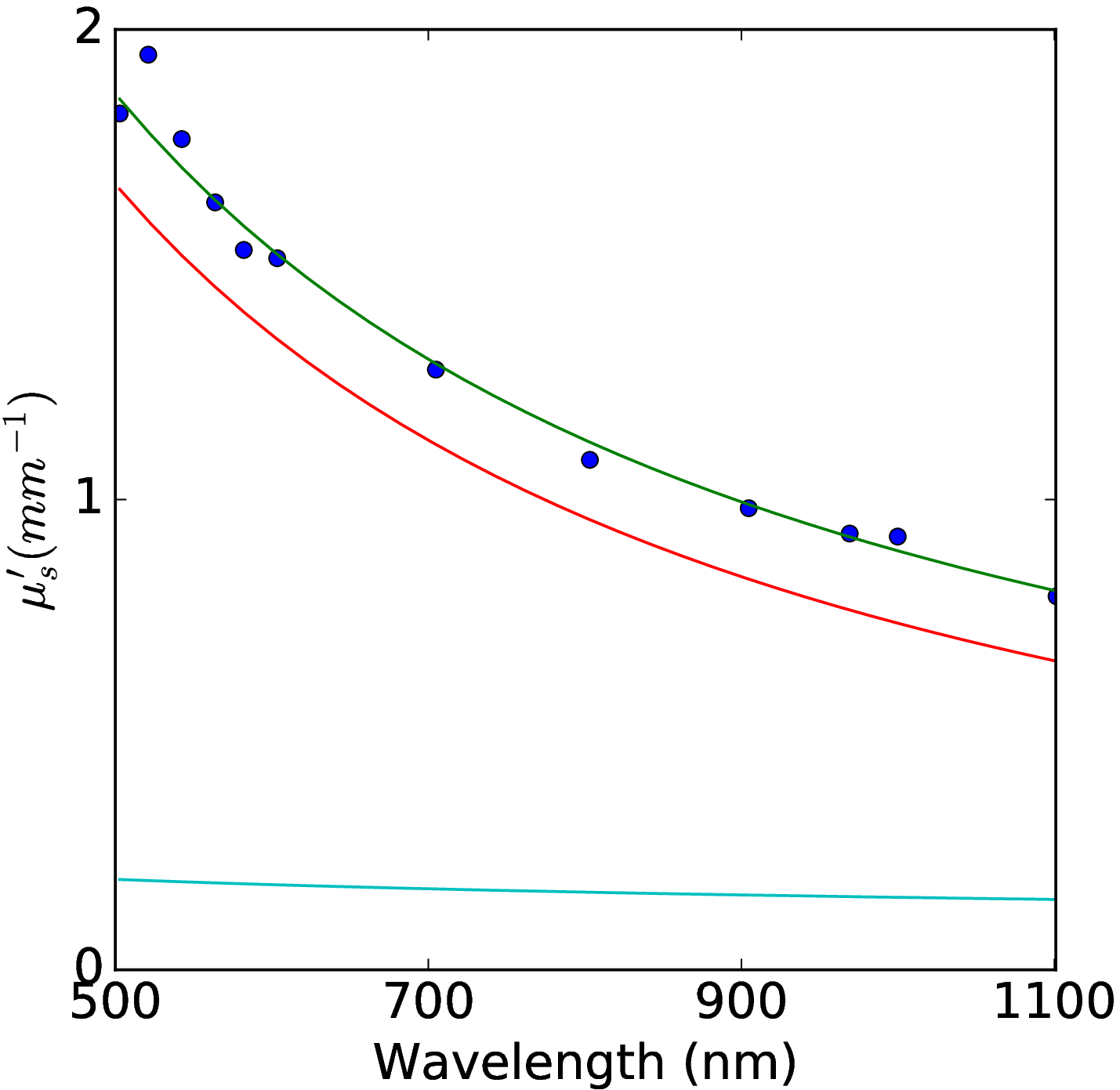}\end{overpic}~\begin{overpic}[width=0.25\columnwidth]{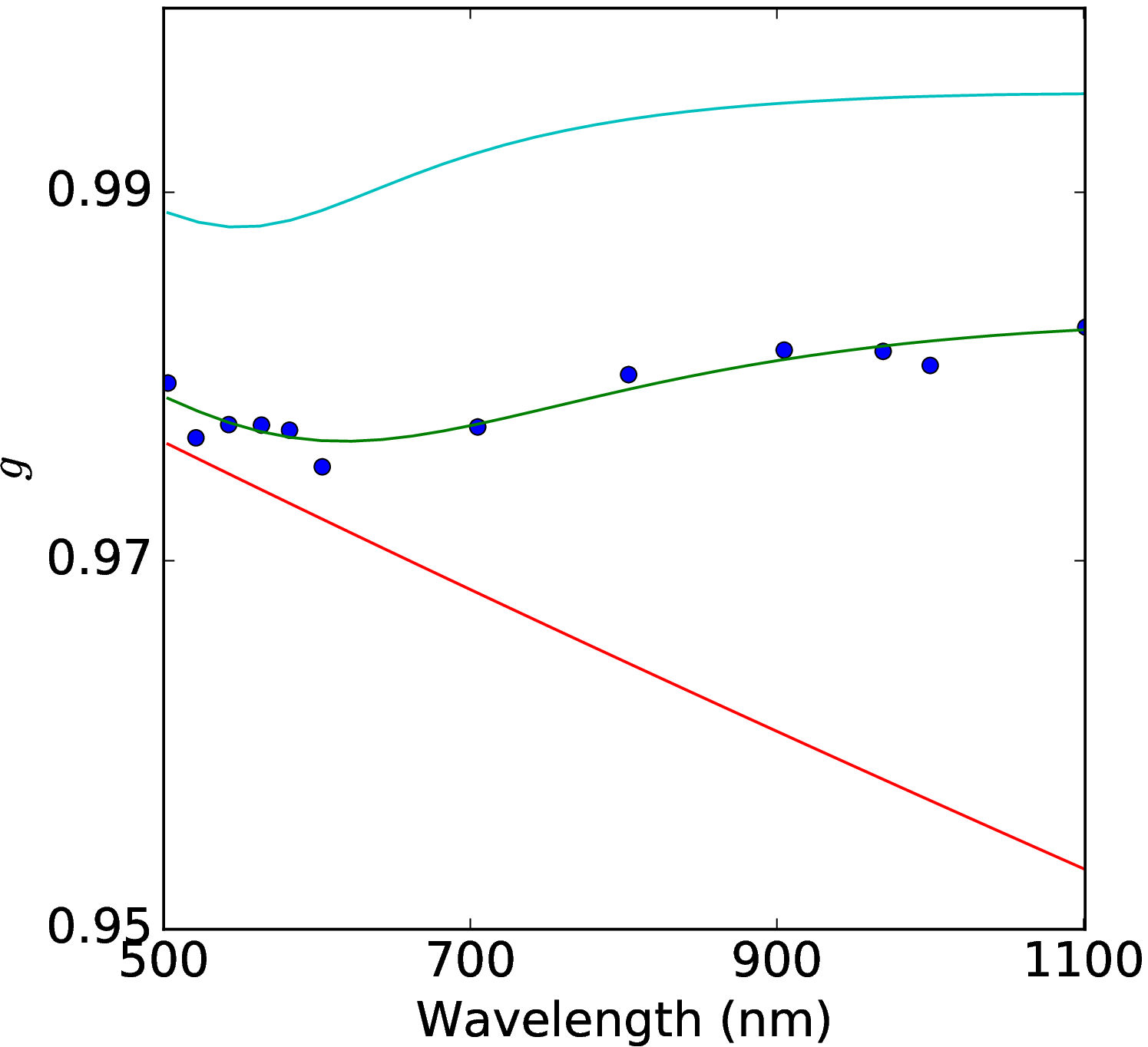}\end{overpic}
\par\end{centering}

\begin{centering}
\begin{overpic}[width=0.25\columnwidth]{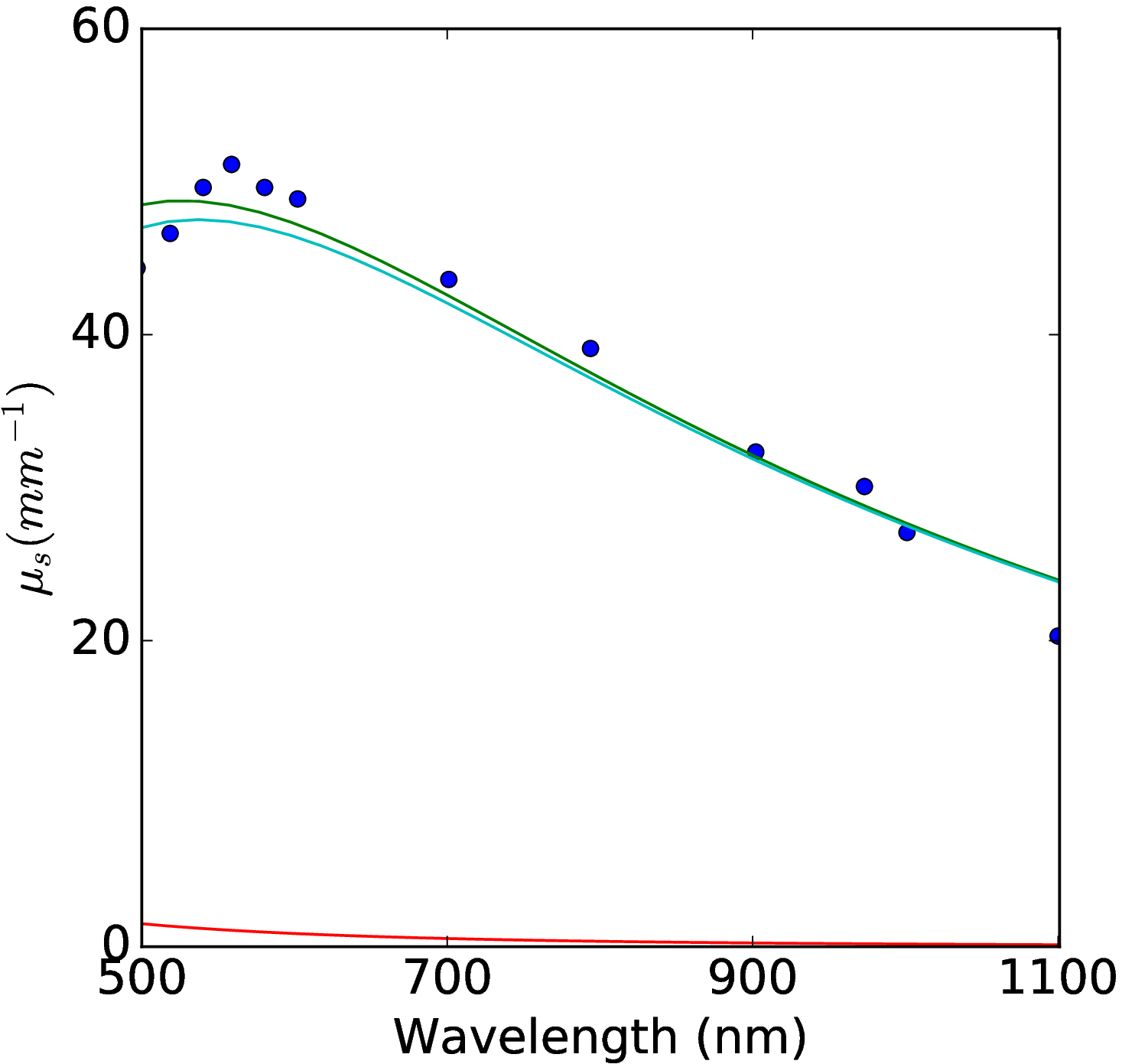}\put(-10,17){\small{(d)}}\end{overpic}~\begin{overpic}[width=0.25\columnwidth]{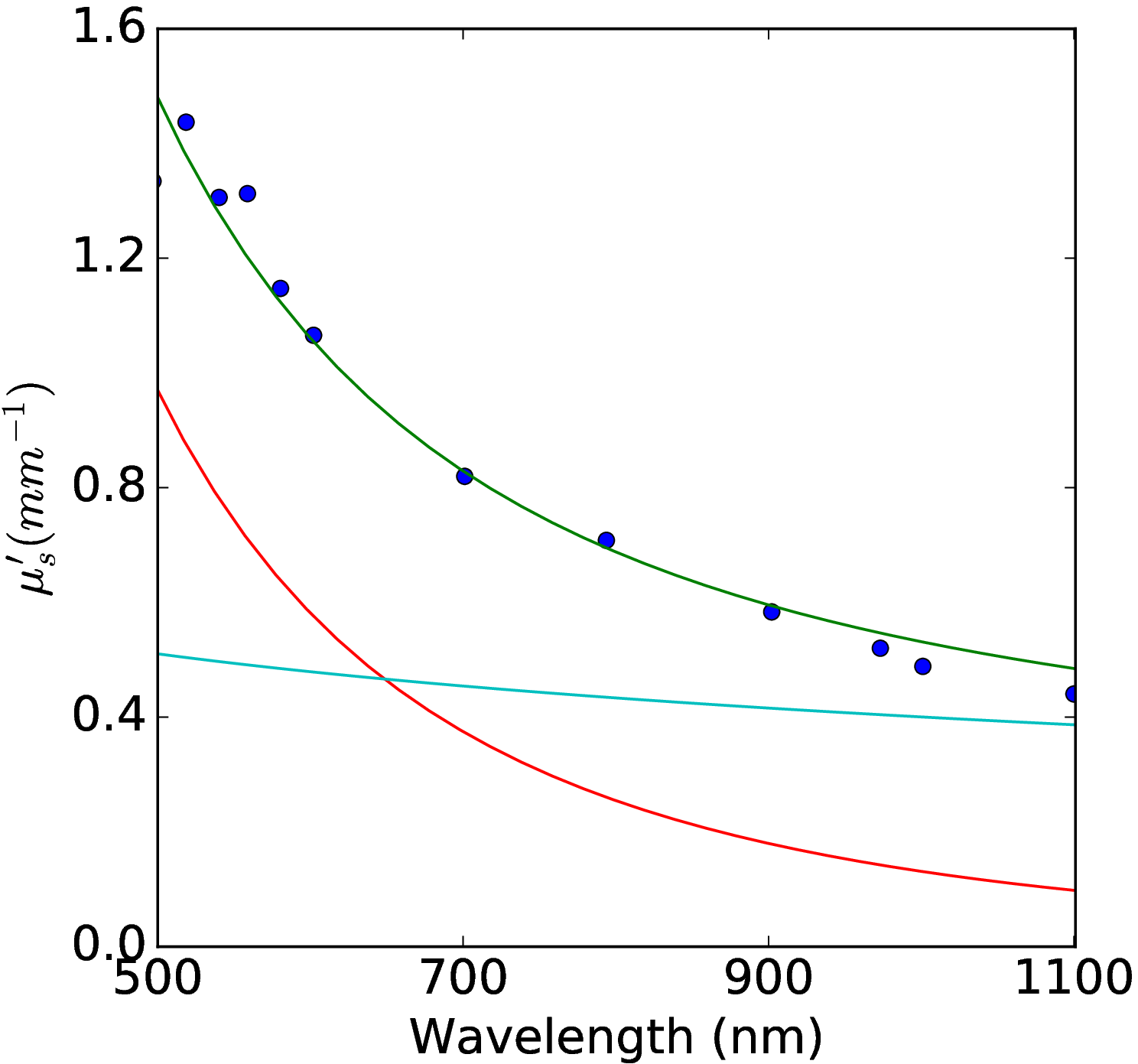}\end{overpic}~\begin{overpic}[width=0.25\columnwidth]{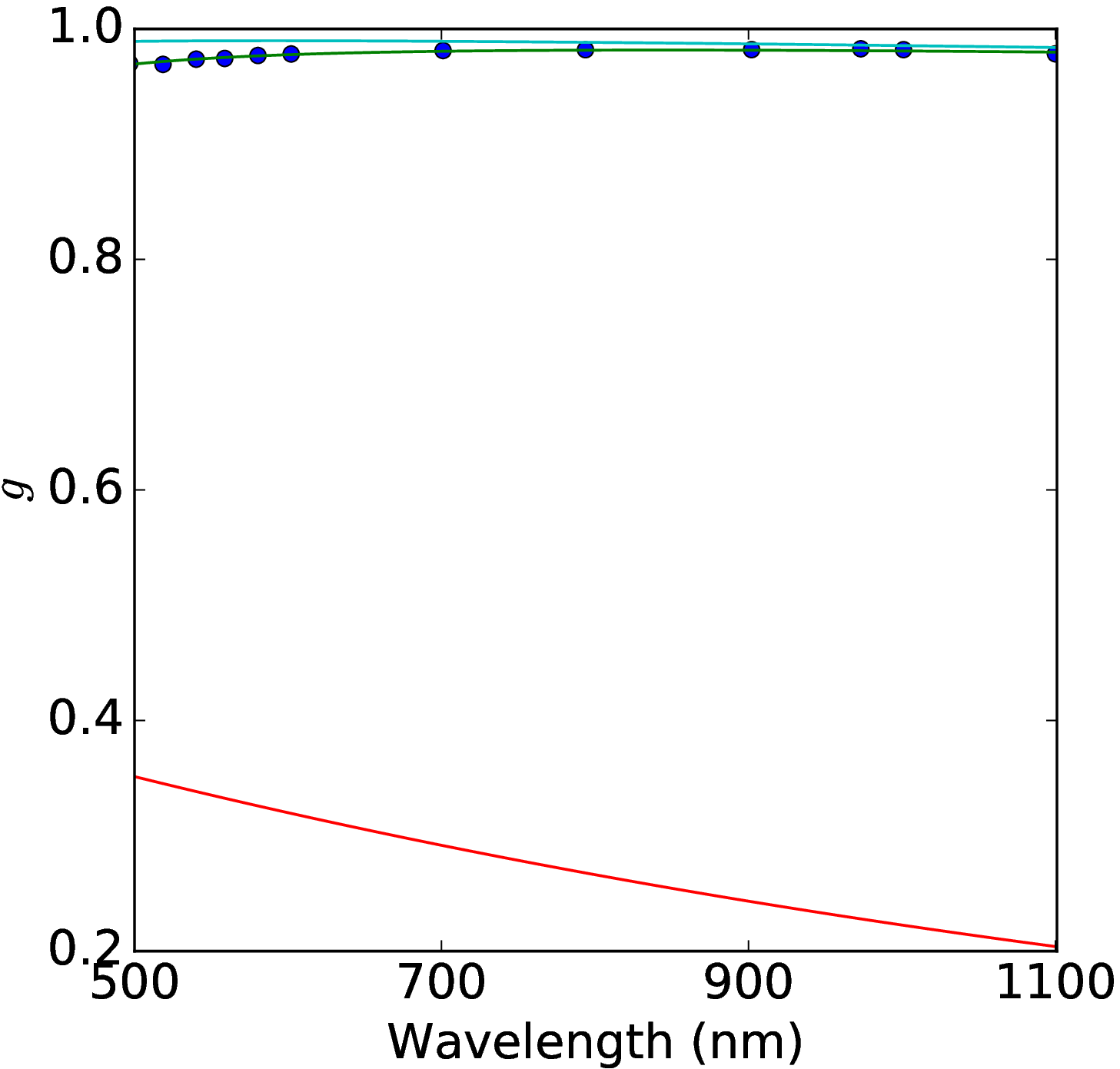}\end{overpic}
\par\end{centering}

\begin{centering}
\begin{overpic}[width=0.25\columnwidth]{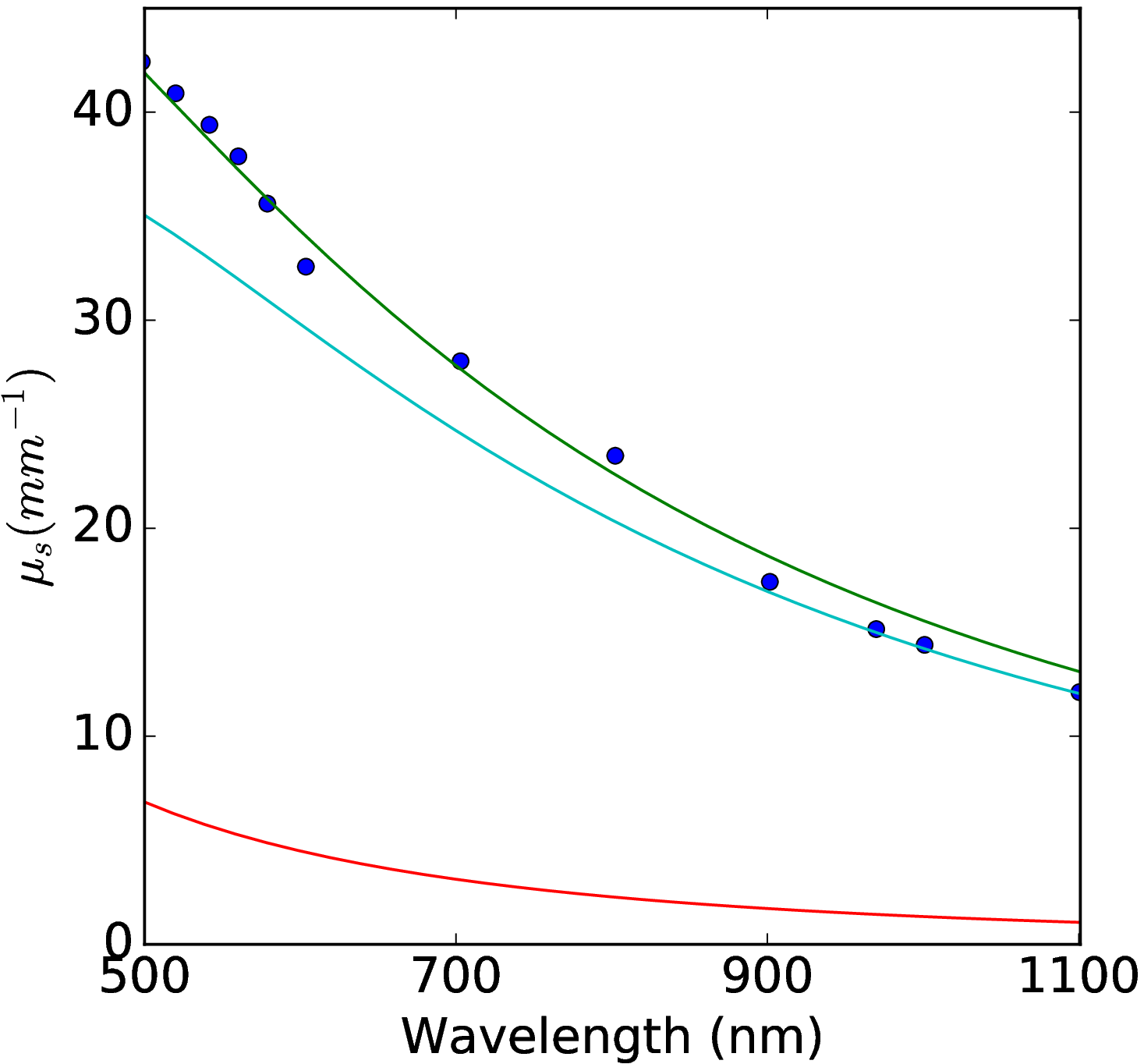}\put(-10,17){\small{(e)}}\end{overpic}~\begin{overpic}[width=0.25\columnwidth]{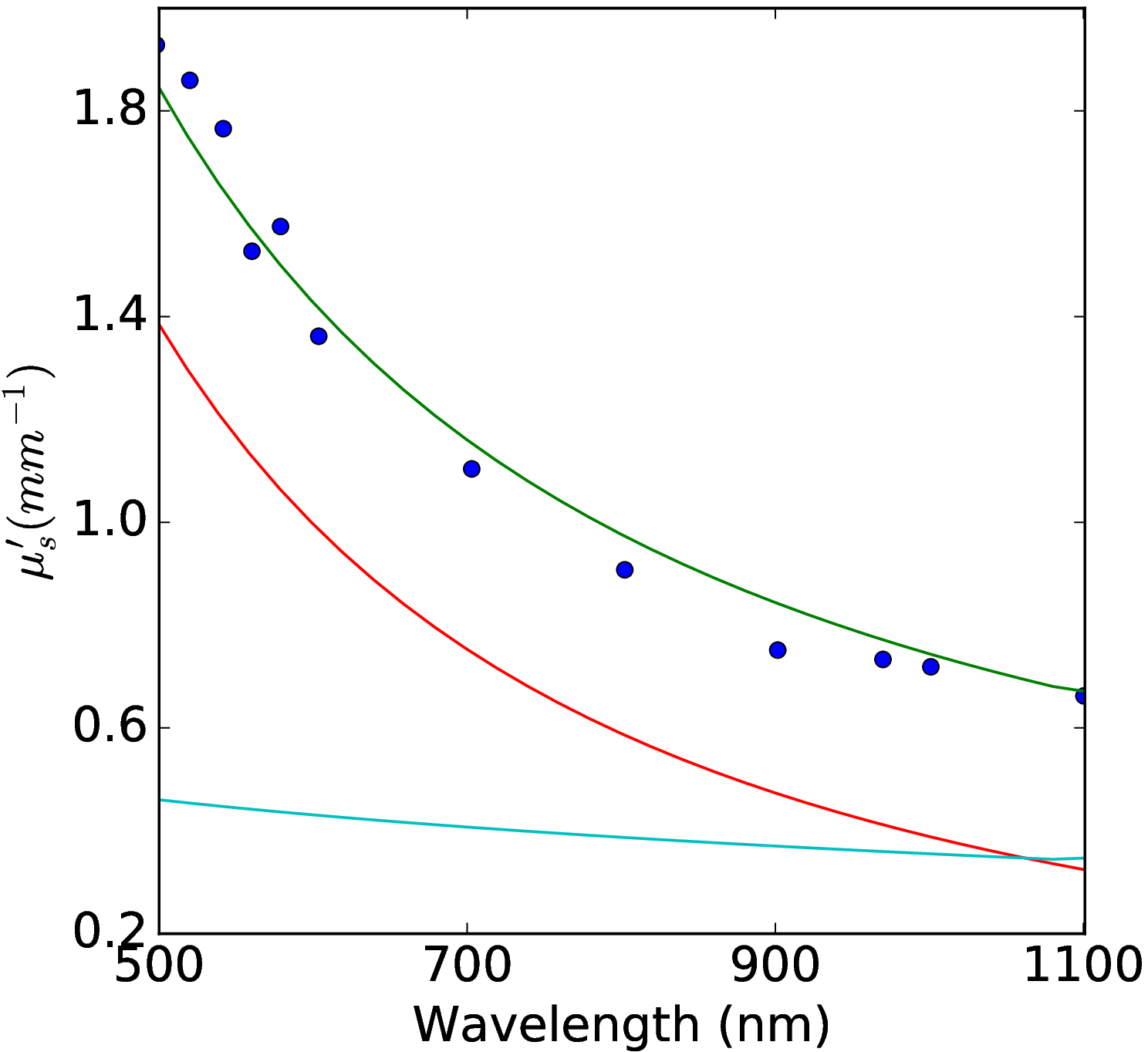}\end{overpic}~\begin{overpic}[width=0.25\columnwidth]{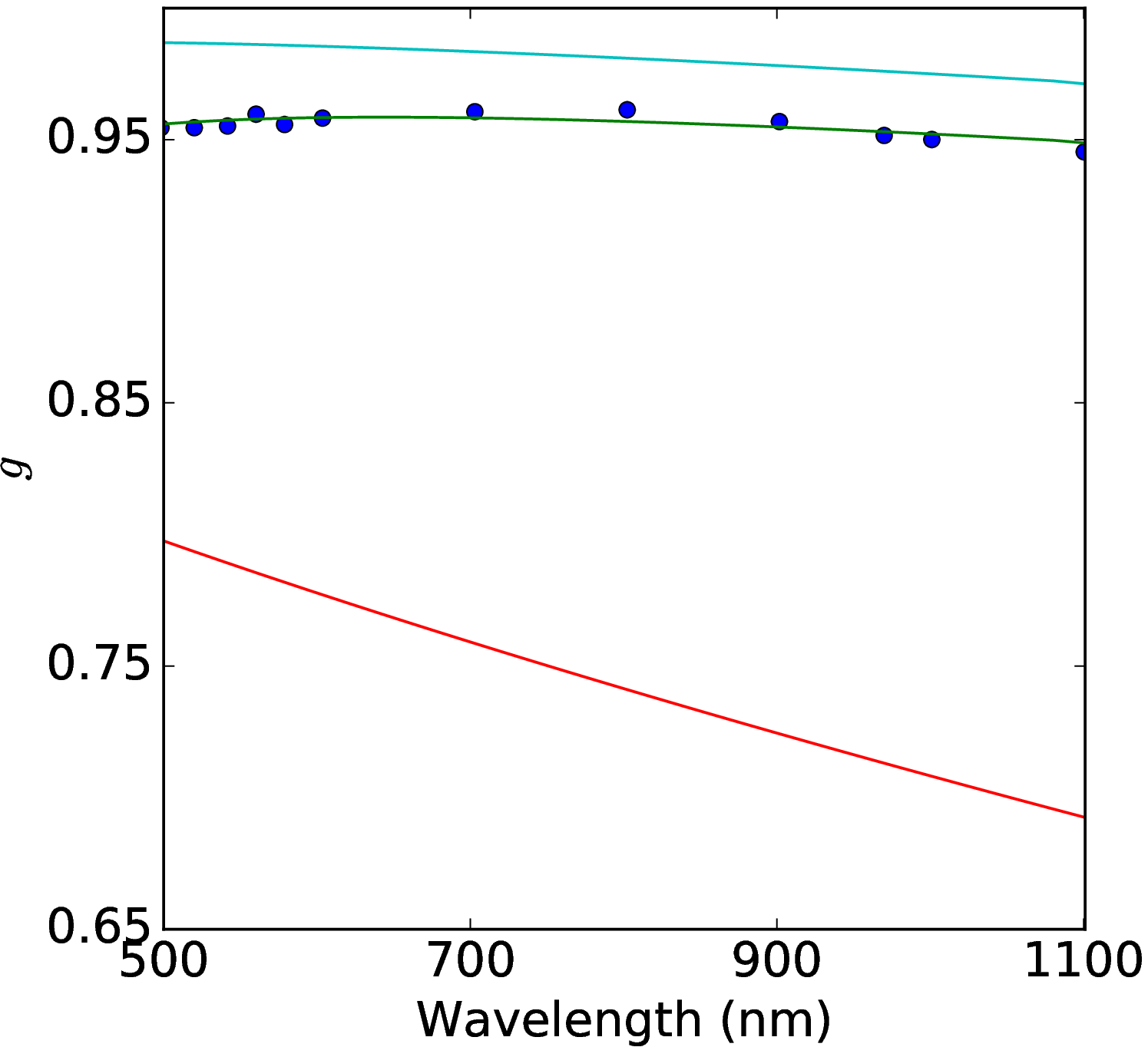}\end{overpic}
\par\end{centering}

\caption{Plum pudding random medium tissue model fitting to from top row to
bottom row (a) normal breast adipose tissue, (b) normal glandular
breast tissue, (c) fibrocystic tissue, (d) fibroadenoma, and (e) ductal
carcinoma. The columns from left to right show $\mu_{s}$, $\mu_{s}'$
and $g$. The background refractive index fluctuation and the core
are shown together with the PPRM tissue model. Experimental data is
adapted from Peters et al \cite{Peters1990}.}

\label{fig:unified tissue fitting breast tissue}
\end{figure}

\begin{figure}
\begin{centering}
\begin{overpic}[width=0.25\columnwidth]{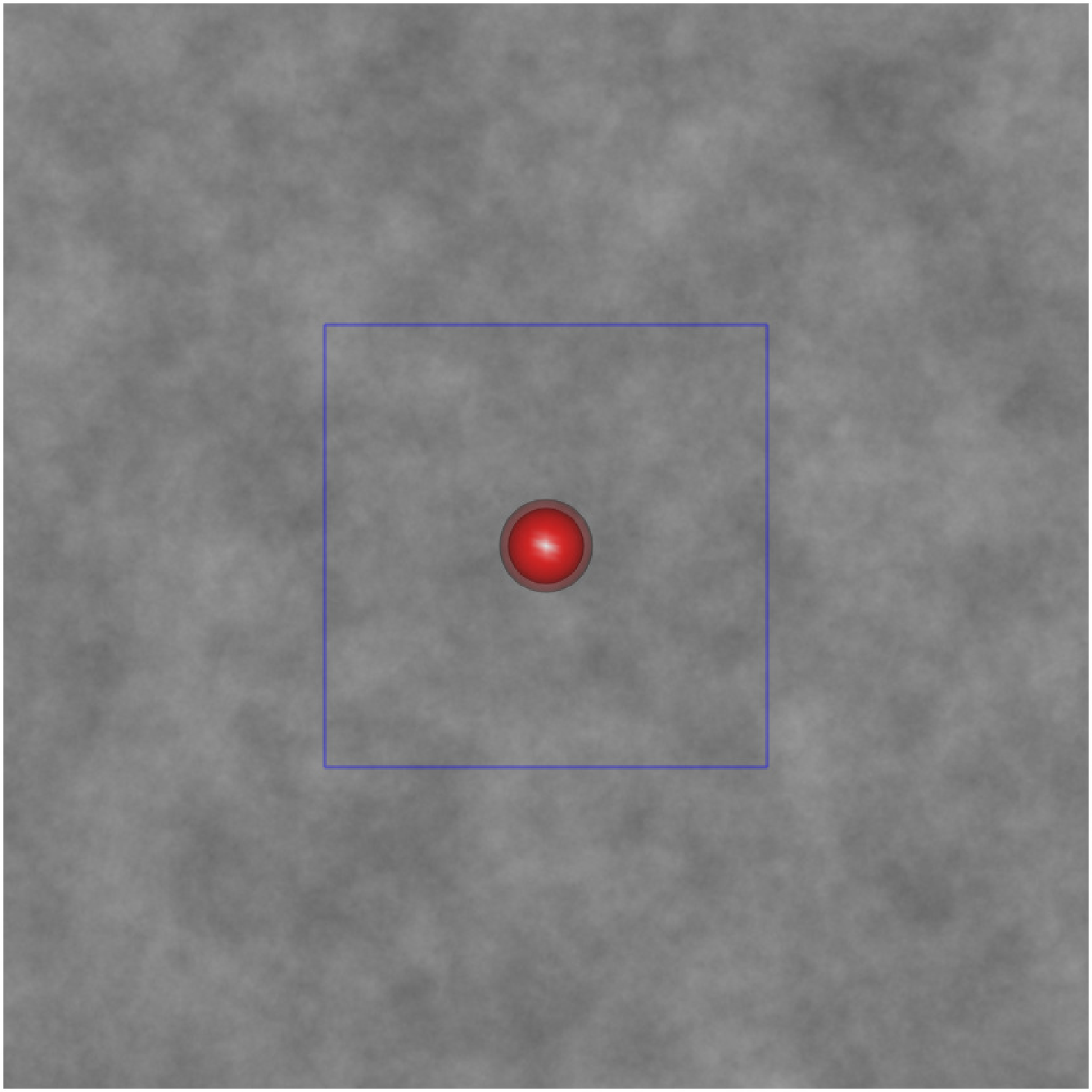}\put(5,85){\color{white} \small{(a)}}\end{overpic}~~\begin{overpic}[width=0.25\columnwidth]{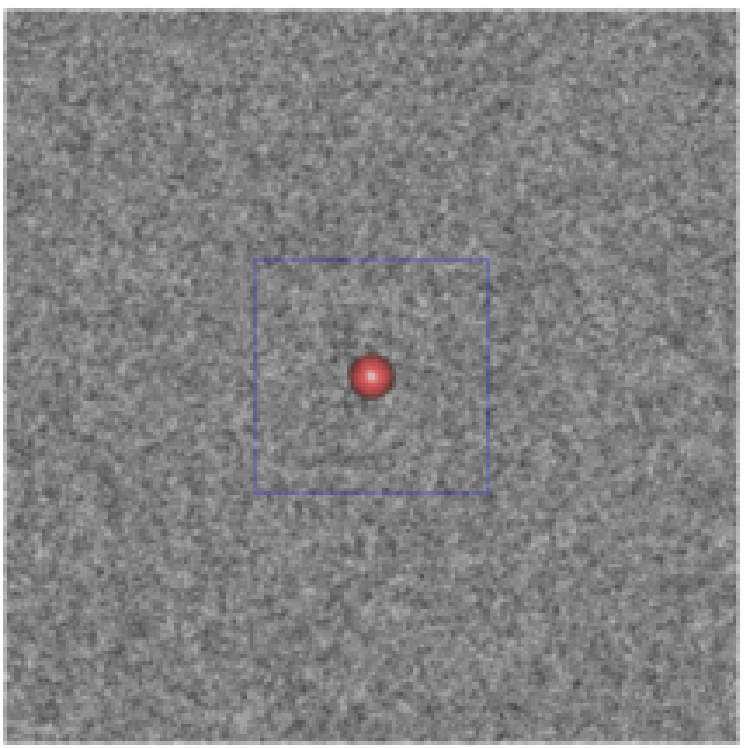}\put(5,85){\color{white} \small{(b)}}\end{overpic}~~\begin{overpic}[width=0.25\columnwidth]{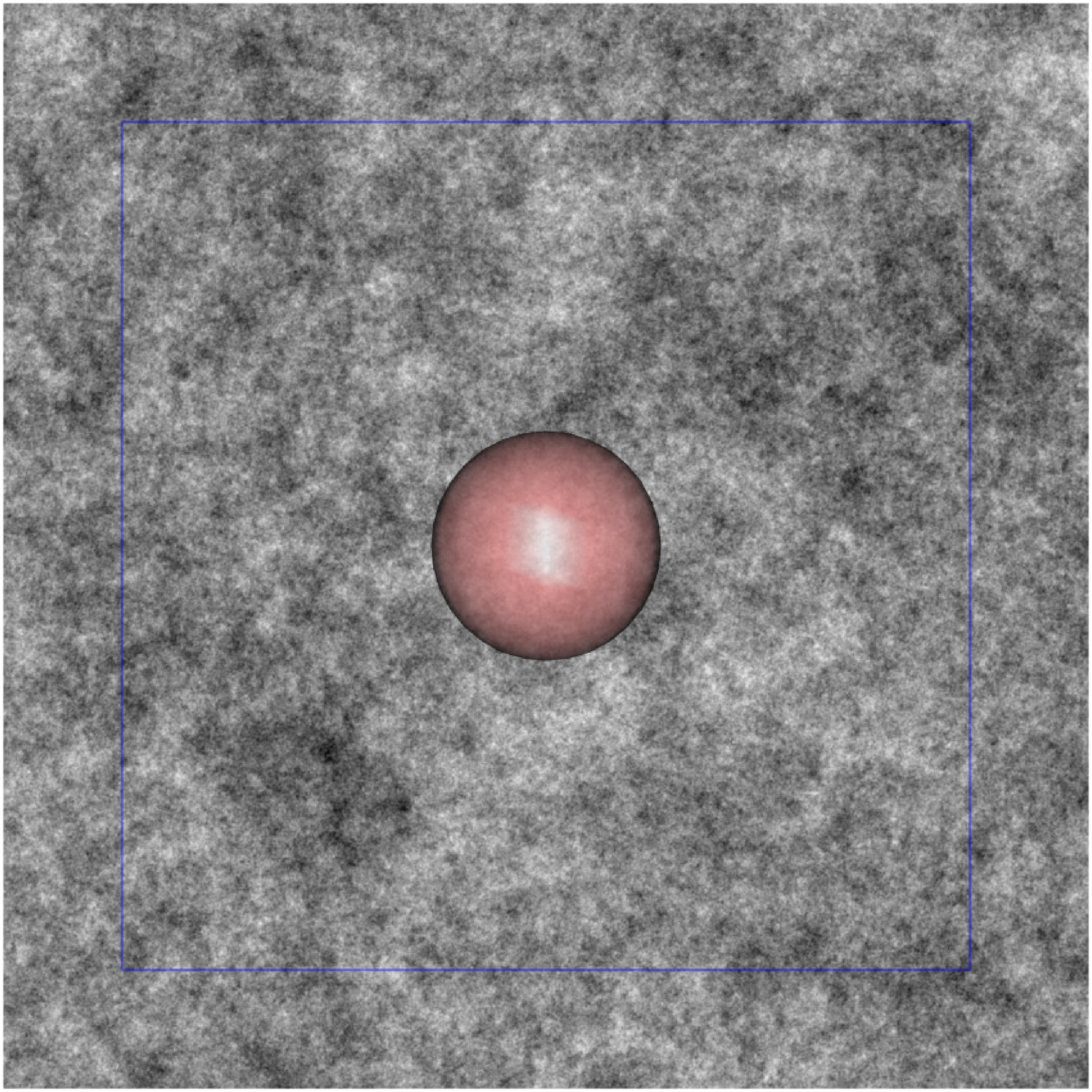}\put(5,85){\color{white} \small{(c)}}\end{overpic}
\par\end{centering}

\vspace{0.25cm}

\begin{centering}
\begin{overpic}[width=0.25\columnwidth]{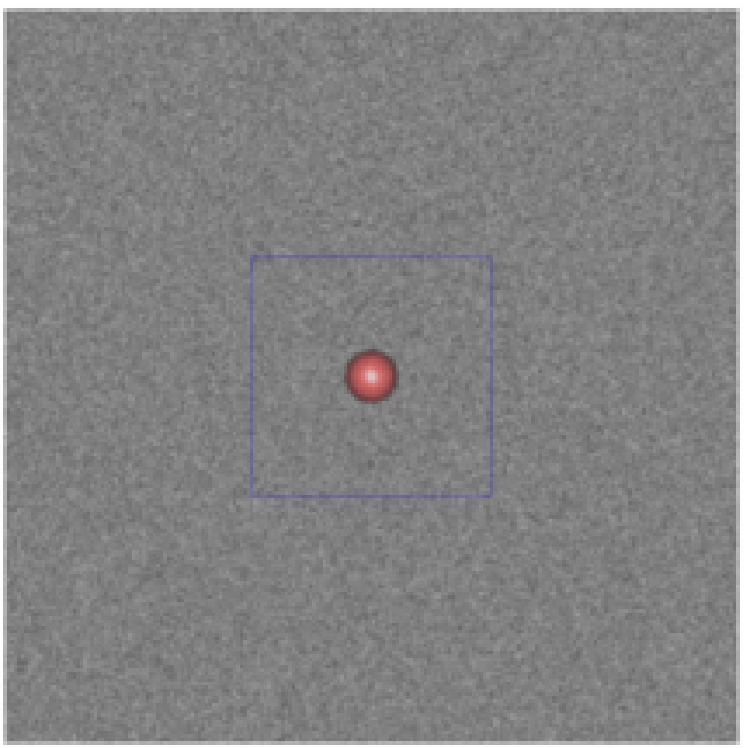}\put(5,85){\color{white} \small{(d)}}\end{overpic}~~\begin{overpic}[width=0.25\columnwidth]{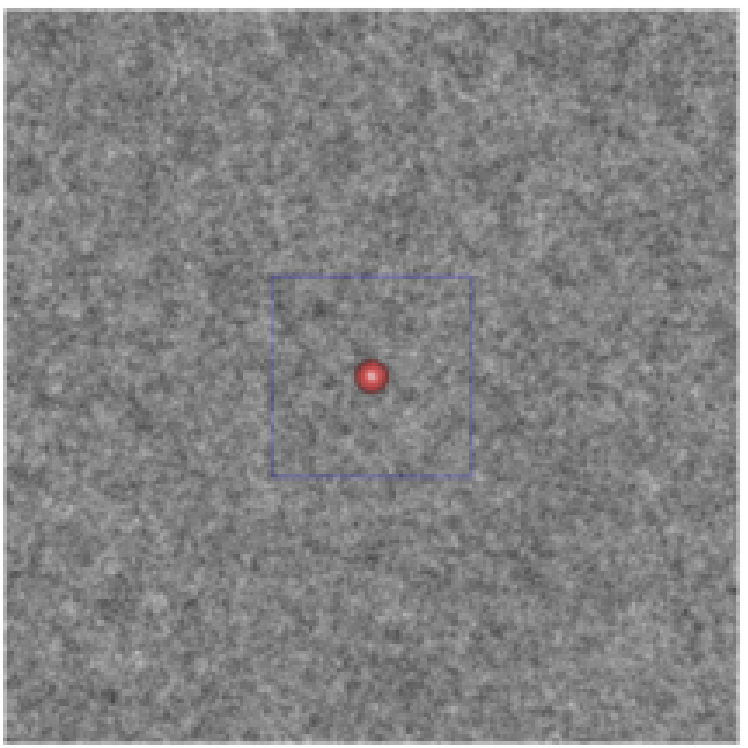}\put(5,85){\color{white} \small{(e)}}\end{overpic}
\par\end{centering}

\vspace{0.25cm}

\begin{centering}
\includegraphics[width=0.31\columnwidth]{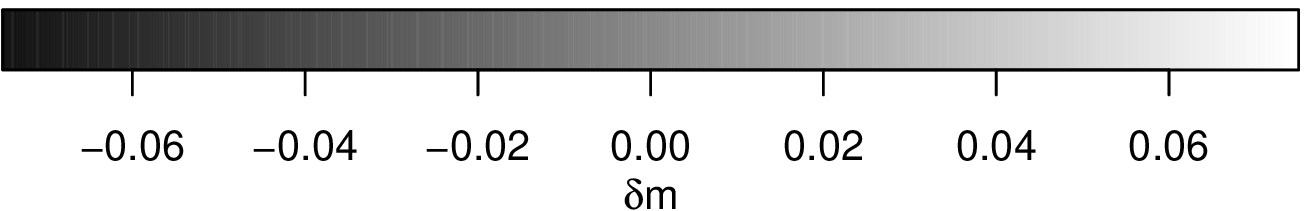}~\includegraphics[width=0.31\columnwidth]{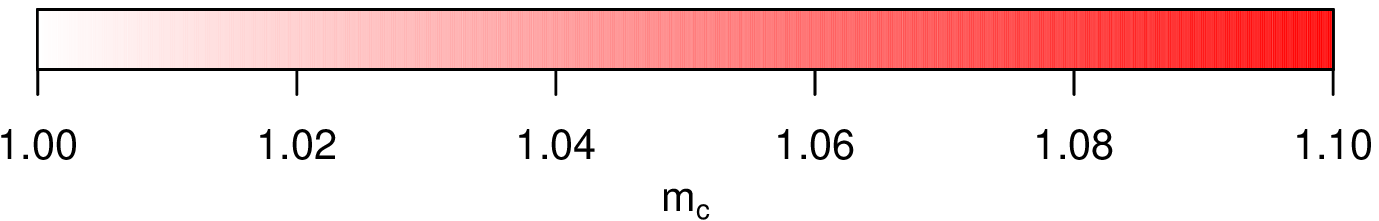}~
\par\end{centering}

\caption{The plum (core) and pudding (background refractive index fluctuation)
in (a) normal breast adipose tissue, (b) normal glandular breast tissue,
(c) fibrocystic tissue, (d) fibroadenoma, and (e) ductal carcinoma.
The whole window size is $30\mu m\times30\mu m$. The blue square
delineates a unit cell which contains exactly one core. A core of
most probable radius $\bar{a}_{c}\exp(-\delta^{2})$ is shown, surrounded
by a shaded area of radius $\bar{a}_{c}\exp(-\delta^{2}+\sqrt{2\log2}\delta)$
at which the number density of the core drops to half maximum. }

\label{fig:core and breast refractive index fluctuation-}
\end{figure}

Most importantly, the Plum Pudding Random Medium tissue model provides
a unified platform to characterize the fine details of structure in
tissue. The microscopic structural alterations in breast tissue associated
with carcinogeneses can be quantified from spectroscopic measurement
alone. Figure \ref{fig:unified tissue fitting breast tissue} shows
the PPRM tissue model fitting to various disease states of freshly
excised and homogenized breast tissue: normal breast adipose tissue,
normal glandular breast tissue, fibrocystic tissue, fibroadenoma,
and ductal carcinoma reported in \cite{Peters1990}. Fibrocystic tissue
and fibroadenoma are the most common benign breast conditions. Their
fitted parameters are summarized in Table \ref{tab:unified tissue model tiffing breast tissue}.
The plum (cores) and pudding (background refractive index fluctuation
$\delta m$) over a whole window of size $30\mu m\times30\mu m$ for
all five types of breast tissue simulated with RandomFields \cite{J.2015}
is shown in Fig.~\ref{fig:core and breast refractive index fluctuation-}.
The blue square delineates a unit cell which contains exactly one
core. A core of most probable radius $\bar{a}_{c}\exp(-\delta^{2})$
is shown, surrounded by a shaded area of radius at which the number
density of the core drops to half maximum. 

Significant structural alterations can be observed from Table \ref{tab:unified tissue model tiffing breast tissue}
and Fig.~\ref{fig:core and breast refractive index fluctuation-}.
The background refractive index of breast tissue is $n_{\mathrm{bg}}\simeq1.36$.
The refractive index for the core can be found to be $n_{\mathrm{core}}=1.49,$
$1.45,$ $1.41$, $1.44$, and $1.46$, respectively, for normal breast
adipose tissue, normal glandular breast tissue, fibrocystic tissue,
fibroadenoma, and ductal carcinoma. The values of $n_{\mathrm{core}}$
agree with the respective refractive index of the nucleus in these
different breast tissues \cite{wang2006,Wang2007,Hajihashemi2012}.
The fibrocystic cell nucleus is more round than that in either normal
or malignant breast cells and has a radius around $6\mu m$ \cite{Nandakumar2012}.
The nucleus of normal and malignant breast cancer cells has more complex
structure with the nucleolar size (radius) ranging from $1\mu m$
to $2\mu m$ \cite{Nandakumar2012,Barbisan1998}. The core can hence
be identified as the nucleus or nucleoli inside a nucleus. The normal
adipose tissue is seen to have the background refractive index fluctuation
of the smallest fractal dimension ($D_{f}=1.56$) and the core with
the largest refractive index ($m_{c}=1.097$) and biggest size variability
($\delta=0.177$). The core in the normal fibrocystic tissue has the
least concentration ($N_{c}=7.76\times10^{-5}\mu m^{-3}$), largest
size ($\bar{a}_{c}=6.345\mu m$), smallest refractive index ($m_{c}=1.039$)
and least size variability ($\delta=0.004$). The normal glandular
breast tissue and fibroadenoma share cores of similar characteristics
whereas the background refractive index fluctuation is seen to be
with a half fluctuation amplitude ($\beta=3.34\times10^{-3}$ vs $6.93\times10^{-3}$),
a half correlation length ($l_{\mathrm{max}}=0.111\mu m$ vs $0.198\mu m$),
and an increase in $D_{f}$ ($5.65$ vs $4.59$) in the latter than
those in the former. Ductal carcinoma is seen to be associated with
the core with the highest concentration ($N_{c}=1.83\times10^{-3}\mu m^{-3}$)
and smallest size ($\bar{a}_{c}=1.39\mu m$). 

Similarly, the relative importance of the two components (background
vs core, or, pudding vs plum) in their contributions to $\mu_{s}$
and $\mu_{s}'$ vary significantly with wavelength (see Table \ref{tab:contribution to mus, musp breast}).
The core dominates in $\mu_{s}$ whereas the background refractive
index fluctuation in $\mu_{s}'$ in general. The importance of the
core increases and that of the background decreases with the probing
wavelength for both $\mu_{s}$ and $\mu_{s}'$. 

\begin{table*}
\caption{Fitted parameters for (a) normal breast adipose tissue, (b) normal
glandular breast tissue, (c) fibrocystic tissue, (d) fibroadenoma,
and (e) ductal carcinoma. The fluctuation amplitude $\sqrt{\left\langle \delta m(0)^{2}\right\rangle }$
is computed from $\beta$ by assuming the inner cutoff for the background
refractive index fluctuations to be $l_{\mathrm{min}}=20\mathrm{nm}$. }

\begin{centering}
\resizebox{\textwidth}{!}{%
\begin{tabular}{cccccccccc}
 & \multicolumn{3}{c}{Background} & \multicolumn{4}{c}{Core} &  & \tabularnewline
 & $\beta(\times10^{-3})$ & $l_{\mathrm{max}}(\mu m)$ & $D_{f}$ & $N_{c}(\mu m^{-3})$ & $\bar{a}_{c}(\mu m)$ & $m_{c}$ & $\delta$ & $\sqrt{\left\langle \delta m(0)^{2}\right\rangle }$ & error\tabularnewline
\hline 
Adipose & $6.08$ & $2.280$ & $1.56$ & $5.46\times10^{-4}$ & $2.176$ & $1.097$ & $0.177$ & $0.0039$ & $0.039$\tabularnewline
Glandular & $6.93$ & $0.198$ & $4.59$ & $1.16\times10^{-3}$ & $1.820$ & $1.069$ & $0.070$ & $0.0153$ & $0.035$\tabularnewline
Fibrocystic & $5.58$ & $3.524$ & $4.12$ & $7.76\times10^{-5}$ & $6.345$ & $1.039$ & $0.004$ & $0.0150$ & $0.020$\tabularnewline
Fibroadenoma & $3.34$ & $0.111$ & $5.65$ & $1.07\times10^{-3}$ & $2.054$ & $1.061$ & $0.077$ & $0.0104$ & $0.035$\tabularnewline
Ductal Carcinoma & $3.46$ & $0.561$ & $4.73$ & $1.83\times10^{-3}$ & $1.390$ & $1.070$ & $0.066$ & $0.0131$ & $0.034$\tabularnewline
 &  &  &  &  &  &  &  &  & \tabularnewline
\end{tabular}}
\par\end{centering}

\label{tab:unified tissue model tiffing breast tissue}
\end{table*}

\begin{table*}
\caption{Relative importance of the background refractive index fluctuation
(pudding) vs the core (plum) to the scattering coefficient $\mu_{s}$
and the reduced scattering coefficient $\mu_{s}'$ at the probing
wavelengths of $500\mathrm{nm}$ and $1100\mathrm{nm}$. }

\begin{centering}
\begin{tabular}{ccccc}
 & \multicolumn{2}{c}{$\mu_{s}$} & \multicolumn{2}{c}{$\mu_{s}'$}\tabularnewline
 & $500\mathrm{nm}$ & $1100\mathrm{nm}$ & $500\mathrm{nm}$ & $1100\mathrm{nm}$\tabularnewline
\hline 
Adipose & $46\%:54\%$ & $10\%:90\%$ & $7\%:93\%$ & $6\%:94\%$\tabularnewline
Glandular & $12\%:88\%$ & $3\%:97\%$ & $78\%:22\%$ & $49\%:51\%$\tabularnewline
Fibrocystic & $80\%:20\%$ & $30\%:70\%$ & $90\%:10\%$ & $81\%:19\%$\tabularnewline
Fibroadenoma & $3\%:97\%$ & $1\%:99\%$ & $66\%:34\%$ & $20\%:80\%$\tabularnewline
Ductal Carcinoma & $16\%:84\%$ & $8\%:92\%$ & $75\%:25\%$ & $48\%:52\%$\tabularnewline
 &  &  &  & \tabularnewline
\end{tabular}
\par\end{centering}

\label{tab:contribution to mus, musp breast}
\end{table*}

Biological tissue has a complex structure. The prominent advantage
of the Plum Pudding Random Medium model is that it provides a succinct
description of the complex structure in terms of a continuous medium
of background refractive index fluctuation (pudding) and distinctive
prominent structures embedded inside (plum) and faithfully
reproduces the observed spectroscopic light scattering properties
($\mu_{s}$, $\mu_{s}'$ and $g$) of biological tissue. Both the pudding
and the plum are essential for tissue light scattering. The reduced
scattering coefficient  $\mu_{s}'$ of tissue is dominated by the
fractal scattering pudding which yields its powerlaw dependence on the
wavelength \cite{xu05:_fract}. Moreover,
PPRM provides a potential resolution to the long lasting puzzle that for most biological tissue
the anisotropy increases and light scattering is more forward directed
with the probing wavelength within visible and near-infrared spectral
range. This ``anomalous'' wavelength dependence of $g$ originates
from the core (plum). 

PPRM offers a novel analytical platform to understand
and interpret light scattering by the complex structures in tissue.
In modeling tissue light scattering, PPRM is much more transparent
and efficient than the current approaches based on computational expensive
FDTD simulations\cite{drezek00:_pulsed_finit_differ_time_domain,brock2006:_fdtd,Starosta2009}
which is unable to model real tissue because of its prohibitive computational
cost. Most importantly, PPRM establishes the inherent connection between
tissue structure characteristics and its light scattering spectroscopy
and opens up a new venue of quantifying
the \emph{microscopic} scattering constituents in tissue from \emph{macroscopic}
probing of a bulk with scattered light. Fine microscopic structural
alterations in tissue associated with cancer or physiological variations
can then be deduced from noninvasive macroscopic light scattering
spectroscopic measurements. The characteristics of the pudding (background
refractive index fluctuation) in tissue has widely been used in early
detection and diagnosis of cancer \cite{xu05:_fract,roy2004,badizadegan2004,Zhu2010a}.
The plum (core) in the PPRM model corresponds to the most prominent
scattering structures of higher refractive index in tissue such as the nucleus or the
nucleolus. The quantification of such cores is hence
of great prognostic value. The smaller nucleolar size has been shown
to be correlated with metastatic cells \cite{Nandakumar2012}. A new
cancer grading system based on {the size
of the nucleolus has also been recently adopted for renal cell carcinoma,
which correlates well with prognosis \cite{Delahunt2013}. }The complete
tissue characterization by PPRM will hence be instrumental in early
detection and diagnosis of tissue diseases including cancer. The deeper
understanding of the nature of tissue light scattering also offers
important insight in optical sensing strategies that a method interrogating
$\mu_{s}$ is preferred to that detecting $\mu_{s}'$ when the core
is the target of interest whereas the method detecting $\mu_{s}'$
is better suited for sensing the background refractive index fluctuations
and a suitable choice of the probing wavelength can significantly
enhance the sensitivity to the target of interest. 

\section{Conclusions}

We have presented here a plum pudding random medium tissue model which
captures the key feature of tissue light scattering structures that
tissue behaves approximately as a continuum (pudding) yet with some
prominent structures (plum) which are distinctive from the background
medium. The background refractive index fluctuation is found to be
well described by the fractal continuous medium. The plum pudding
random medium model faithfully reproduces the wavelength dependence
of tissue light scattering and its anisotropy. It provides a potential
resolution to the lasting
puzzle that tissue is scattered more into the forward directions by
light of longer wavelengths and attributes the ``anomalous'' trend
in the anisotropy of light scattering to the plum whereas the pudding
gives rise to the powerlaw dependence of the reduced scattering
coefficient on the probing wavelength. Most importantly,
the plum pudding random medium model opens up a novel venue of remote
sensing of tissue architecture and microscopic structures from
spectroscopic light scattering. With a complete quantification of the plum and the pudding
of tissue, PPRM accurately depicts fine tissue structural alterations\emph{
on average} associated with tissue disease states or physiological
variations without excising the tissue. 

One prominent advantage of optical methods is the rich spectroscopic
content in tissue-light interactions and the potential to probe morphological,
biochemical and functional structure of tissue noninvasively. The
plum pudding random medium tissue model establishes the quantitative
connection between the rich spectroscopic content in light scattering
and the underlying tissue microstructure. PPRM may find wide applications
in understanding and modeling tissue light scattering, and enabling
remote microscopy from spectroscopic scattered light, for example,
the promising development in \emph{in vivo} optical histopathology
of deep tissue from spectroscopic diffuse light measurement. 

\appendix
\def\erf{\mathop{\operator@font erf}\nolimits}
\def\arctanh{\mathop{\operator@font arctanh}\nolimits}
\def\arccsc{\mathop{\operator@font arccsc}\nolimits}
\def\sinc{\mathop{\operator@font sinc}\nolimits}
\def\Si{\mathop{\operator@font Si}\nolimits}

\section*{Appendix A: The large size limit for the fractal and Whittle-Matern continuum
model} \label{sec:appendix-large-size-limit}

See Table \ref{tab:random  medium large size limit}.

\begin{table*}
\caption{The light scattering expressions of the fractal and Whittle-Matern
continuous medium model in the limit of $X=kl_{\mathrm{max}},kl\gg1$
where $\alpha\equiv\left\langle \delta m(0)^{2}\right\rangle \pi^{1/2}\frac{\Gamma(\nu+3/2)}{\left|\Gamma(\nu)\right|}$,
and $X=2\pi n_{0}l_{\mathrm{max}}/\lambda$ and $2\pi n_{0}l/\lambda$,
respectively, in the fractal and Whittle-Matern model. Both models
behave alike in this limit and their scattering properties dependence
on wavelength reduces to a power law with an identical power if interchanging
$D_{f}$ and $(4-2\nu)$.}

\begin{centering}
\smaller{%
  \begin{tabular}{ccc}
 & Fractal Model & Whittle-Matern Model\tabularnewline
\hline 
$\mu_{s}$ & %
\begin{minipage}[t]{0.48\textwidth}%
\begin{eqnarray*}
\frac{2}{5-D_{f}}\beta^{2}l_{\mathrm{max}}^{-1}X^{2}\\
(D_{f}<5)\\
\frac{2^{D_{f}-5}(23-8D_{f}+D_{f}^{2})}{(D_{f}-1)(D_{f}-3)}\frac{\pi}{\sin\left(\frac{D_{f}-5}{2}\pi\right)}\beta^{2}l_{\mathrm{max}}^{-1}X{}^{D_{f}-3}\\
(5<D_{f}<7)
\end{eqnarray*}
\end{minipage} & %
\begin{minipage}[t]{0.45\textwidth}%
\begin{eqnarray*}
\frac{4\alpha}{1+2\nu}l^{-1}X^{2}\\
(\nu>-\frac{1}{2})\\
\alpha\frac{(7+4\nu^{2})}{(3-2\nu)(4\nu^{2}-1)}2^{1-2\nu}l^{-1}X^{1-2\nu}\\
(-\frac{1}{2}>\nu>-\frac{3}{2})
\end{eqnarray*}
\end{minipage}\tabularnewline
$\mu_{s}'$ & %
\begin{minipage}[t]{0.48\textwidth}%
\begin{eqnarray*}
2\frac{D_{f}-4+(3-D_{f})\ln(2X)}{(3-D_{f})^{2}}\beta^{2}l_{\mathrm{max}}^{-1}\\
(D_{f}<3)\\
  \frac{2^{D_{f}-3}(D_{f}^{2}-4D_{f}+11)}{(D_{f}-3)(D_{f}^{2}-1)}\frac{\frac{5-D_{f}}{2}\pi}{\sin\left(\frac{5-D_{f}}{2}\pi\right)}\beta^{2}l_{\mathrm{max}}^{-1}X^{D_{f}-3}\\
(3<D_{f}<7)
\end{eqnarray*}
\end{minipage} & %
\begin{minipage}[t]{0.45\textwidth}%
\begin{eqnarray*}
\frac{4\alpha}{4\nu^{2}-1}l^{-1}\\
(\nu>\frac{1}{2})\\
\alpha\frac{(4\nu^{2}-8\nu+11)}{(1-2\nu)(3-2\nu)(5-2\nu)}2^{2-2\nu}l^{-1}X^{1-2\nu}\\
(\frac{1}{2}>\nu>-\frac{3}{2})
\end{eqnarray*}
\end{minipage}\tabularnewline
$1-g$ & %
\begin{minipage}[t]{0.48\textwidth}%
\begin{eqnarray*}
(5-D_{f})\frac{D_{f}-4+(3-D_{f})\ln(2X)}{(3-D_{f})^{2}}X^{-2}\\
(D_{f}<3)\\
\frac{2^{D_{f}-5}(D_{f}^{2}-4D_{f}+11)}{(D_{f}-3)(D_{f}^{2}-1)}\frac{(5-D_{f})^{2}\pi}{\sin\left(\frac{5-D_{f}}{2}\pi\right)}X^{D_{f}-5}\\
(3<D_{f}<5)\\
2\frac{(D_{f}-5)(D_{f}^{2}-4D_{f}+11)}{(D_{f}+1)(23-8D_{f}+D_{f}^{2})}\\
(5<D_{f}<7)
\end{eqnarray*}
\end{minipage} & %
\begin{minipage}[t]{0.45\textwidth}%
\begin{eqnarray*}
\frac{1}{2\nu-1}X^{-2}\\
(\nu>\frac{1}{2})\\
\frac{(1+2\nu)(4\nu^{2}-8\nu+11)}{(1-2\nu)(3-2\nu)(5-2\nu)}2^{-2\nu}X^{-1-2\nu}\\
(\frac{1}{2}>\nu>-\frac{1}{2})\\
-2\frac{(2\nu+1)(4\nu^{2}-8\nu+11)}{(5-2\nu)(7+4\nu^{2})}\\
(-\frac{1}{2}>\nu>-\frac{3}{2})
\end{eqnarray*}
\end{minipage}\tabularnewline
\end{tabular}}
\par\end{centering}

\label{tab:random  medium large size limit}
\end{table*}

\section*{Appendix B: Lognormal size distribution of the core}\label{appendix:logn-size-distr}

The radius of the polydisperse core is assumed to follow a lognormal
distribution, 
\begin{equation}
f(a)=\frac{1}{\sqrt{2\pi}\delta}a^{-1}\exp\left[-\ln^{2}(\frac{a}{\bar{a}})/2\delta^{2}\right].
\end{equation}
The lognormal size distribution of parameters $\bar{a}$ and $\delta$
attains its peak at $\bar{a}/\exp(\delta^{2})$ and a full width at
half maximum (FWHM) of the size distribution to be $2\sinh(\sqrt{2\ln2}\delta)\bar{a}/\exp(\delta^{2})$.
The two important characteristics of the size distribution are the
effective radius 
\begin{equation}
a^{\mathrm{eff}}=\frac{\int_{0}^{\infty}a^{3}f(a)da}{\int_{0}^{\infty}a^{2}f(a)da}=a\exp(5\delta^{2}/2)
\end{equation}
and the effective variance 
\begin{equation}
\nu^{\mathrm{eff}}=\frac{\int_{0}^{\infty}(a-a^{\mathrm{eff}})^{2}a^{2}f(a)da}{(a^{\mathrm{eff}})^{2}\int_{0}^{\infty}a^{2}f(a)da}=\exp(\delta^{2})-1.
\end{equation}
These two characteristics are geometrical projection area weighted.
Scatterers of different size distribution but of the same effective
radius and variance behave alike in their properties of light scattering
\cite{mishchenko02:_scatt}. 

\section*{Appendix C: Empirical expressions for light scattering efficiencies of an optically
soft particle}\label{appendix:empir-expr-light}

The scattering cross section for the core is given by 
\begin{equation}
C_{\mathrm{sca}}(x,m)=\pi a^{2}Q_{\mathrm{sca}}(x,m)\label{eq:Csca large soft particle}
\end{equation}
where $x\equiv ka$ is the size parameter with $a$ being the radius
of the core and $Q_{\mathrm{sca}}$ is the scattering efficiency.
In analog to Eq.~(\ref{eq:Csca large soft particle}), a similar efficiency
can be defined for the reduced scattering cross section $C_{\mathrm{sca}}'$
weighted by $1-\mu$, i.e., 
\begin{equation}
C_{\mathrm{sca}}'(x,m)=\pi a^{2}Q_{\mathrm{sca}}'(x,m)
\end{equation}

The scattering efficiency is very well described by the anomalous
diffraction theory for optically soft particles \cite{hulst81:_light}
\begin{equation}
Q_{\mathrm{sca}}(x,m)=2-\frac{4}{\eta}\sin\eta+\frac{4}{\eta^{2}}(1-\cos\eta),\label{eq:Q Mie}
\end{equation}
where $\eta\equiv2x(m-1)$ is the optical delay for a ray passing
through the center of the particle. There are, however, no simple
analytical expressions for $Q_{\mathrm{sca}}'$, $Q_{\mathrm{sca}}''$
and $\gamma\equiv Q_{\mathrm{sca}}''/Q_{\mathrm{sca}}'$ where $Q_{\mathrm{sca}}''$
is defined in a similar fashion to $Q_{\mathrm{sca}}'$ with the weighting
factor $(1-\cos\theta)$ replaced by $(1-P_{2}(\cos\theta))$ where
$P_{2}$ is the second order Legendre polynomial. For a size parameter
$10<x<200$ and $\left|m-1\right|\le0.05$, simple empirical expressions
can be fitted from the exact Mie solution as following:\begingroup\medmuskip=0mu\thickmuskip=0mu

\begin{equation}
Q_{\mathrm{sca}}'(x,m)=2\pi\left|m-1\right|^{2}(0.578-3.256\left|m-1\right|)x^{2.690\left|m-1\right|+0.217},
\end{equation}
\begin{equation}
Q_{\mathrm{sca}}''(x,m)=\gamma(x,m)Q_{\mathrm{sca}}'(x,m)
\end{equation}
\endgroup and\begingroup\medmuskip=0mu\thickmuskip=0mu
\begin{eqnarray}
\gamma(x,m) & = & 32.2\left|m-1\right|^{2}-5.22\left|m-1\right|+1.929
  -(0.1528\left|m-1\right|-0.00076) \\
 &  & \times\frac{\left(\eta-12.097+424.72\left|m-1\right|-4465.1\left|m-1\right|^{2}\right)}{\sqrt{\eta}}.\nonumber 
\end{eqnarray}
\endgroup The average relative error within the regime is $0.85\%$
and $0.37\%$ for $Q_{\mathrm{sca}}'$ and $\gamma$, respectively.
Their maximum relative error does not exceed $6\%$ and $2\%$. The
anisotropy factor is given by $g(x,m)=1-Q_{\mathrm{sca}}'/Q_{\mathrm{sca}}$.
The empirical expressions (in solid lines) and the exact values from
Mie theory (in symbols) for $Q_{\mathrm{sca}}$, $Q_{\mathrm{sca}}'$,
$g$ and $\gamma$ are displayed in Fig.~\ref{fig:Qsca', gamma soft particle}. 

\begin{figure}
\begin{centering}
~~\includegraphics[width=0.3\columnwidth]{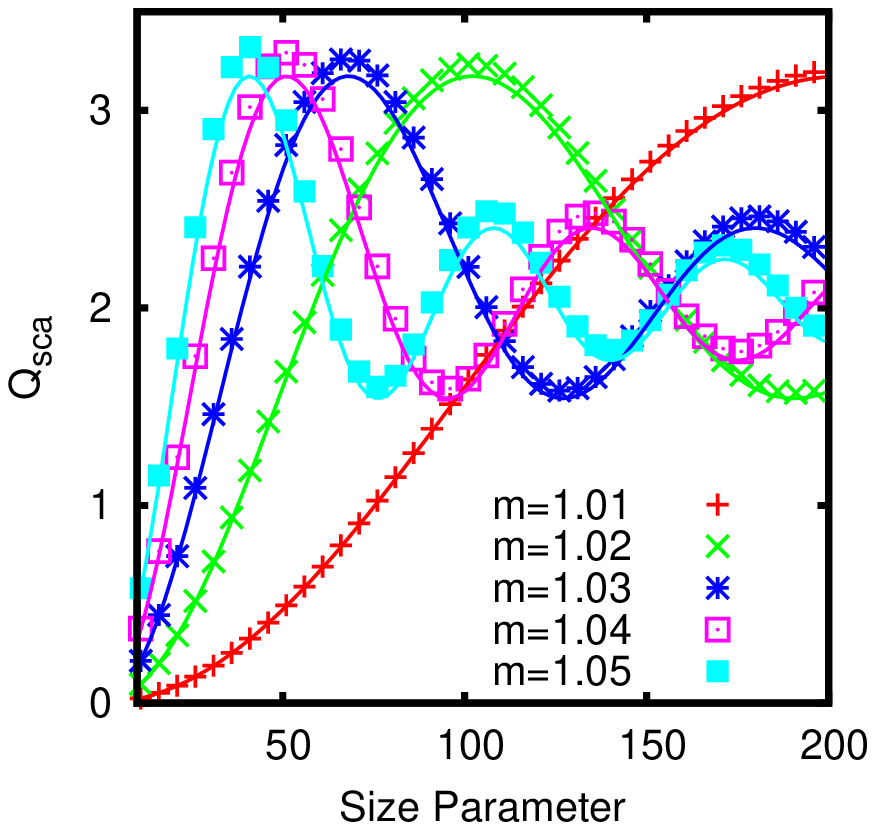}~~~\includegraphics[width=0.3\columnwidth]{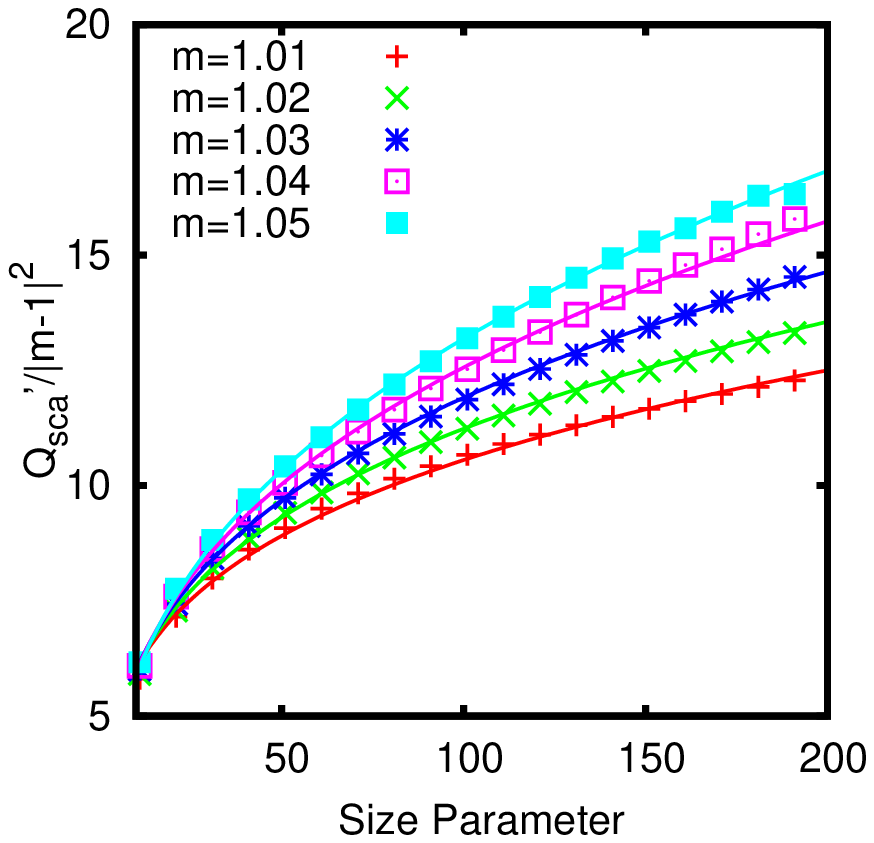}
\par\end{centering}

\begin{centering}
\includegraphics[width=0.32\columnwidth]{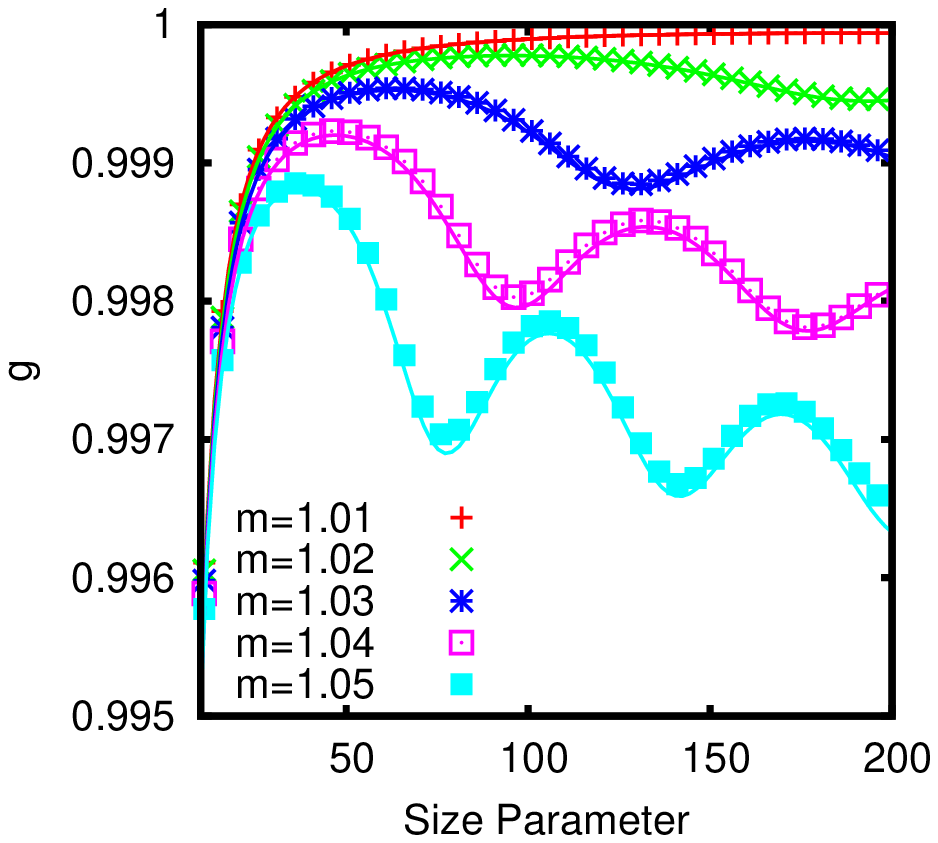}~~~\includegraphics[width=0.3\columnwidth]{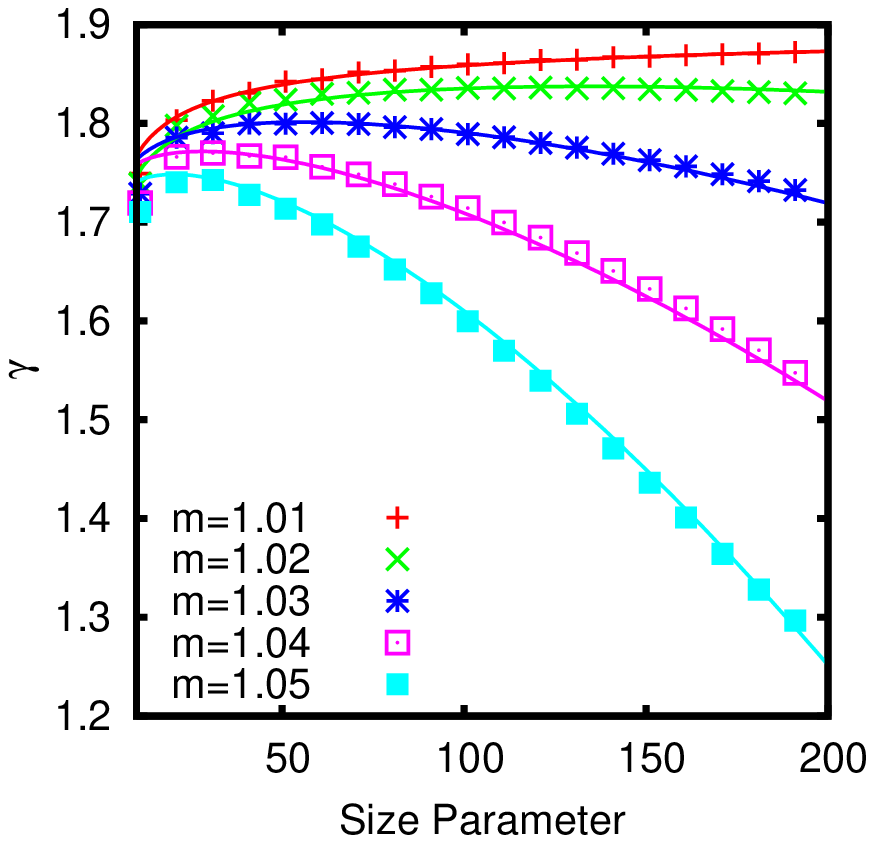}
\par\end{centering}

\caption{The scattering efficiency $Q_{\mathrm{sca}}$, the reduced scattering
efficiency $Q_{\mathrm{sca}}'$, the anisotropy factor $g$, and the
$\gamma$ factor for optically soft particle of size parameter $10<x<200$.
The empirical expressions (in solid lines) and the exact values from
Mie theory (in symbols) are shown for $m=1.01$, $m=1.02$, ..., $m=1.05$. }

\label{fig:Qsca', gamma soft particle}
\end{figure}

\begin{figure}
\begin{centering}
\includegraphics[width=0.3\columnwidth]{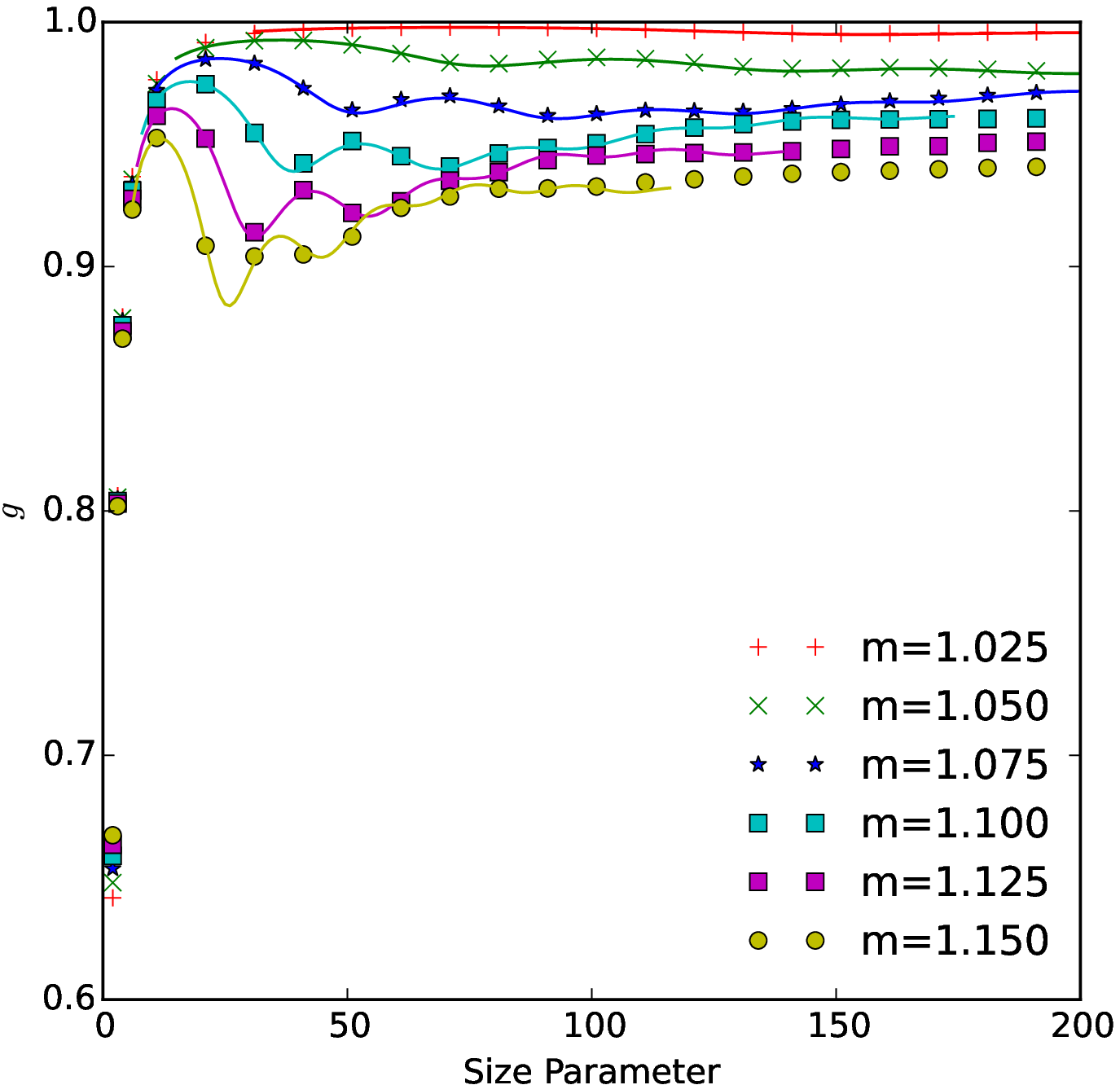}~~~\includegraphics[width=0.3\columnwidth]{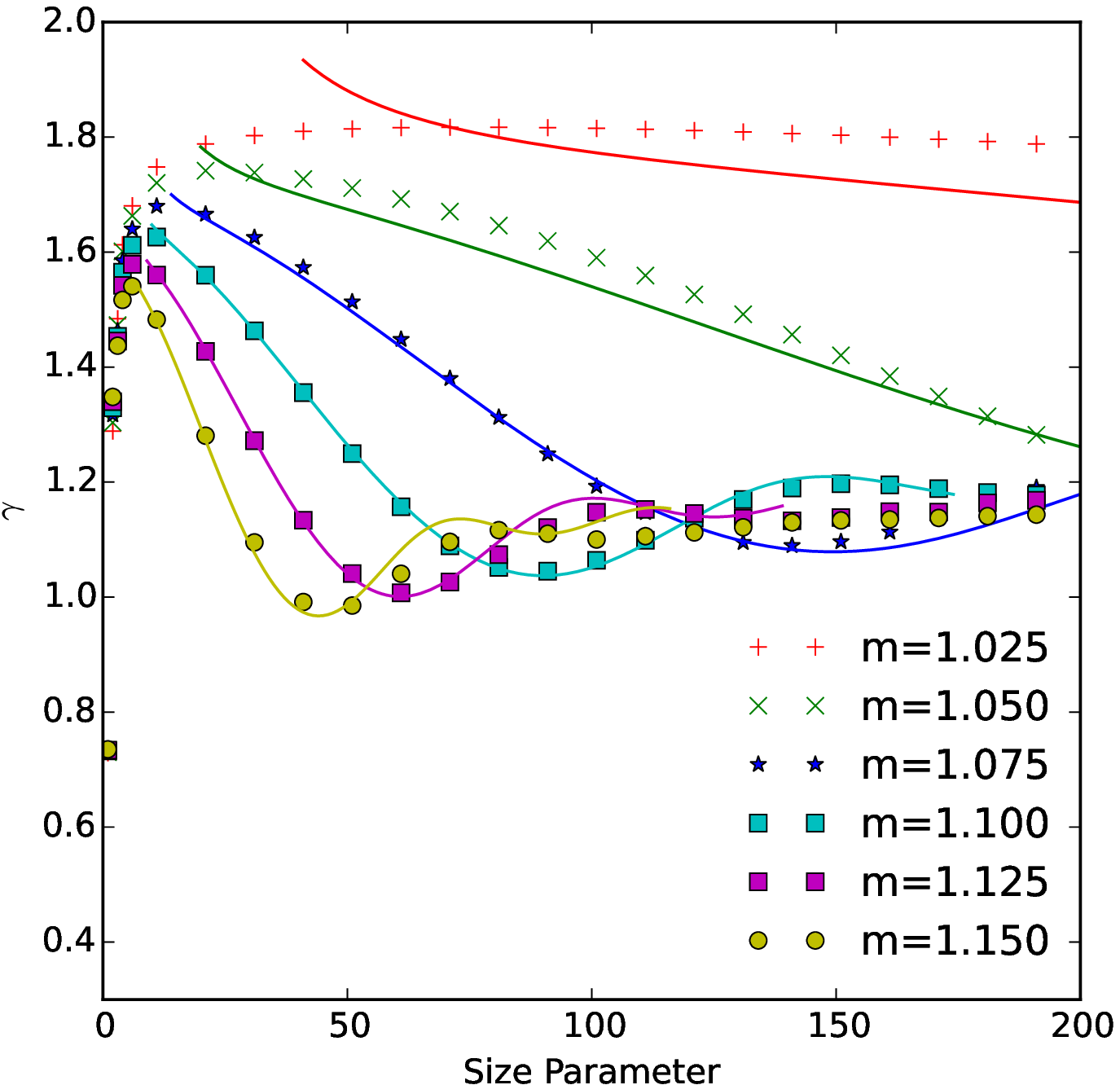}~
\par\end{centering}

\caption{The anisotropy factor $g$ and $\gamma$ for polydispersed soft particles
($\nu^{\mathrm{eff}}=1.0\%$). The symbols are from Mie calculations
and the solid lines are from the empirical expressions for $g=1-\bar{Q}_{\mathrm{sca}}'/\bar{Q}_{\mathrm{sca}}$
and $\gamma=\bar{Q}_{\mathrm{sca}}''/\bar{Q}_{\mathrm{sca}}'$ over
the range $1\le\left|m-1\right|x\le17.5$.}

\label{fig:polydisperse g and g2}
\end{figure}

For polydisperse soft particles following a lognormal size distribution,
empirical expressions have also been obtained over the region $\left|m-1\right|\le0.15$
and $0<x<200$ for $\bar{Q}_{\mathrm{sca}}$, $\bar{Q}_{\mathrm{sca}}'$
and $\bar{Q}_{\mathrm{sca}}''$ in the form of
\resizebox{\textwidth}{!}{%
  $2\pi\left(1+d_{1}\left|m-1\right|\right)\left|m-1\right|^{d_{2}} 
  \left[c_{0}+c_{1}-(c_{1}+c_{3})\frac{\sin(c_{4}y)}{c_{4}y}\exp(-c_{2}y)\right.
  \left.+2c_{3}\frac{1-\cos(c_{6}y)\exp(-c_{5}y)}{(c_{6}y)^{2}}\right]$}
where $y\equiv x\left|m-1\right|^{d_{3}}$. The parameters are given
in Table \ref{tab:paramerts for Qsca bar etc}. The polydispersity
of the scatterers averages out the highly oscillatory terms in Mie
scattering and hence extends the valid region for the empirical expressions.
The value given by the empirical expression for $\bar{Q}_{\mathrm{sca}}$
has a mean squared root error of $2.4\%$ and the maximum error less
than $5.0\%$ as long as $\left|m-1\right|x\ge1$. When $1\le\left|m-1\right|x\le15$,
the value given by the empirical expressions for $\bar{Q}_{\mathrm{sca}}'$
has a mean squared root error of $1.8\%$ and the maximum error less
than $5.5\%$ whereas the value given by the empirical expression
for $\bar{Q}_{\mathrm{sca}}''$ has a mean squared root error of $1.5\%$
and the maximum error less than $6.9\%$. The anisotropy factor $g$
and $\gamma$ for polydispersed soft particles are plotted in Fig.~\ref{fig:polydisperse g and g2}. 

\begin{table*}
\caption{Parameters for empirical expressions of $\bar{Q}_{\mathrm{sca}}$,
$\bar{Q}_{\mathrm{sca}}'$ and $\bar{Q}_{\mathrm{sca}}''$.}

\begin{centering}
\begin{tabular}{ccccccccccc}
 & $c_{0}$ & $c_{1}$ & $c_{2}$ & $c_{3}$ & $c_{4}$ & $c_{5}$ & $c_{6}$ & $d_{1}$ & $d_{2}$ & $d_{3}$\tabularnewline
\hline 
$\bar{Q}_{\mathrm{sca}}$ & 0 & 0.3161 & 0.0768 & 0.2204 & 2.0465 & -0.0103 & 1.5213 & 0.712 & 0 & 1\tabularnewline
$\bar{Q}_{\mathrm{sca}}'$ & 0.4292 & 0.1704 & 0.3908 & 1.0386 & 2.2673 & -0.0300 & 2.6834 & -1.095 & 1.613 & 1.725\tabularnewline
$\bar{Q}_{\mathrm{sca}}''$ & 0.6572 & 0.0465 & 0.6964 & 1.4930 & 3.1693 & -0.0358 & 4.0711 & -1.642 & 1.579 & 1.806\tabularnewline
\end{tabular}
\par\end{centering}

\label{tab:paramerts for Qsca bar etc}
\end{table*}

\section*{Appendix D: Fitting procedure}\label{appendix:fitting-procedure}

As the reduced scattering is often dominated by the background refractive
index fluctuation, we adopt the following multi-step procedure when
we fit PPRM to the measured spectroscopic data: (1) Fit the fractal
or Whittle-Matern continuum model alone to the observed reduced scattering
coefficient data for $(l_{\mathrm{max}},D_{f})$ or $(l,\nu)$; (2)
Fit the PPRM tissue model to the observed $g(\lambda)$ for $\bar{a}_{c}$,
$m_{c}$,$\delta$, and $N_{c}$ by fixing the parameters obtained
in (1) unchanged; and (3) Fit the PPRM tissue model to all observed
data $\mu_{s}'(\lambda)$, $\mu_{s}(\lambda)$ and $g(\lambda)$ using
the results from (1) and (2) as the initial guess. To avoid trapping
inside a local minimum, global minimization with basin hopping \cite{Wales1997}
or the particle swarm algorithm is used in the last step.

The fitting error reported in Table 1 and 2 is the mean least squared
error defined by
\begin{eqnarray*}
  \left[\bar{\mu}_{s,\mathrm{meas}}^{-2}\sum_{\lambda}(\mu_{s,\mathrm{mod}}-\mu_{s,\mathrm{meas}})^{2}\right.
  +\bar{\mu}_{s,\mathrm{meas}}'^{-2}\sum_{\lambda}(\mu_{s,\mathrm{mod}}'-\mu_{s,\mathrm{meas}}')^{2}
  \left.+\bar{g}_{\mathrm{meas}}^{-2}\sum_{\lambda}(g_{\mathrm{mod}}-g_{\mathrm{meas}})^{2}\right]^{1/2}
\end{eqnarray*}
between the model (``mod'') and the measurement (``meas'') where
$\bar{\mu}_{s,\mathrm{meas}}$ etc are the average of the measured
data to homogenize the contributions from $\mu_{s}$, $\mu_{s}'$
and $g$. 

\section*{Funding }
NIH (1R15EB009224); DOD (W81XWH-10-1-0526).

\section*{Acknowledgements}
This work was supported by (1R15EB009224) and DOD
(W81XWH-10-1-0526). We thank Swapan K Gayen and Al Katz of City
College of New York for their useful comments.

\end{document}